\documentclass[aps,pra,twocolumn,amsmath,amssymb,footinbib,showpacs]{revtex4}

\usepackage[english]{babel}
\usepackage{latexsym}
\usepackage{graphics}
\usepackage{graphicx}
\usepackage{subfigure}
\usepackage{epsfig}
\usepackage{exscale}
\usepackage{amsmath}
\usepackage{amsfonts}
\usepackage{amssymb}
\usepackage{amscd}
\usepackage{bm}           
\usepackage{bbm}
\usepackage{color}
\usepackage{ulem}
\usepackage{mathrsfs}

\newcommand{\BE}{\begin{equation}}
\newcommand{\EE}{\end{equation}}
\newcommand{\be}{\begin{equation}}
\newcommand{\ee}{\end{equation}}
\newcommand{\beq}{\begin{eqnarray}}
\newcommand{\eeq}{\end{eqnarray}}
\newcommand{\kommentar}[1]{}

\newcommand{\bc}{\begin{center}}
\newcommand{\ec}{\end{center}}

\newcommand{\ud}{\mathrm{d}}

\newcommand{\vecr}{\textbf{r}}
\newcommand{\vecp}{\textbf{p}}

\newcommand{\Vr}{V_{\textrm{\tiny R}}}
\newcommand{\sigmar}{\sigma_{\textrm{\tiny R}}}
\newcommand{\Er}{E_{\textrm{\tiny R}}}
\newcommand{\taur}{\tau_{\textrm{\tiny R}}}
\newcommand{\pr}{p_{\textrm{\tiny R}}}
\newcommand{\Vtrap}{V_{\textrm{\tiny trap}}}
\newcommand{\ntrap}{n_{\textrm{\tiny trap}}}

\newcommand{\me}{\mathrm{e}}

\begin{document} 

\title{Regimes of classical transport of cold gases in a two-dimensional anisotropic disorder}

\author{L.~Pezz\'e, 
M.~Robert-de-Saint-Vincent\footnote{Current address: Physikalisches Institut der Universit\"at Heidelberg Philosophenweg 12, 69120 Heidelberg, Germany}, 
T.~Bourdel, 
J.-P.~Brantut\footnote{Current address: Institute for Quantum Electronics, ETH Z\"urich, H\"unggerberg, CH-8093 Z\"urich, Switzerland}, 
B.~Allard, 
T.~Plisson, 
A.~Aspect, 
P.~Bouyer, 
and L.~Sanchez-Palencia}

\address{Laboratoire Charles Fabry de l'Institut d'Optique,
CNRS and Univ. Paris-Sud, campus Polytechnique RD128,
F-91127 Palaiseau cedex, France}

\begin{abstract}
We numerically study the dynamics of cold atoms in a two-dimensional disordered potential.
We consider an anisotropic speckle potential and focus on the classical dynamics,
which is relevant to some recent experiments.
Firstly, we study the behavior of particles with a fixed energy
and identify different transport regimes.
For low energy, the particles are classically localized due 
to the absence of a percolating cluster.
For high energy, the particles undergo normal diffusion
and we show that the diffusion coefficients 
scale algebraically with the particle energy,
with an anisotropy factor which significantly differs from that of the disordered potential.
For intermediate energy, we find a transient sub-diffusive regime,
which is relevant to the time scale of typical experiments.
Secondly, we study the behavior of a cold-atomic gas with an arbitrary 
energy distribution, using the above results as a groundwork.
We show that the density profile of the atomic cloud
in the diffusion regime is strongly peaked and, in particular,
that it is not Gaussian. 
Its behavior at large distances allows us to extract the energy-dependent 
diffusion coefficients from experimental density distributions.
For a thermal cloud released into the disordered potential, we show that
our numerical predictions are in agreement with experimental findings.
Not only does this work give insights to recent experimental results,
but it {may also serve} interpretation of future experiments searching for deviation from
classical diffusion and traces of Anderson localization.
\end{abstract}

\date{\today}


\maketitle
\tableofcontents

\section{Introduction}
\label{introduction}

Transport processes are widespread and
common in many fields of physics, chemistry 
and biology and have
drawn the attention of both theorists and experimentalists.
Diffusion of waves and particles in disordered
media determines the behaviors of a variety of physical systems. 
These include electronic conductivity in dirty materials \cite{Lee_1985},
superfluidity of liquid helium in porous substrates \cite{Reppy_1992},
and diffusion of light in dense media \cite{Akkermans_book},
with applications to dielectric materials \cite{John_1991}
or interstellar clouds \cite{Chandrasekhar_book}.
Diffusion is at the heart of transport phenomena in
most systems of condensed-matter physics.
For instance, it underlies the Drude theory of
conductivity \cite{Ashcroft_book}, as well as the
self-consistent theory of Anderson localization \cite{Vollhardt_1981}.
A major hindrance to the complete understanding of condensed-matter systems
is the complex interplay of many relevant ingredients, such as
the structure and thermal fluctuations of the underlying substrate \cite{Ashcroft_book},
the disorder \cite{Anderson_1958}, and the inter-particle interactions,
which can lead either to superconductivity \cite{deGennes_book}
or to metal-insulator transitions \cite{Mott_1968,Mott_1990}.
This poses a number of fundamental questions
regarding the existence and the nature of disorder-induced metal-insulator transitions \cite{Vollhardt_1992},
the role of the spatial dimension and the effect of non-linearities.

{Currently, much attention} is devoted to disordered quantum gases
\cite{Fallani_2008,Modugno_2010,Sanchez-Palencia_2010, Lagendijk_2009, Aspect_2009}.
These systems offer unprecedented control of most parameters,
including temperature, geometry and particle-particle interaction
\cite{Dalfovo_1999,Giorgini_2008,Lewenstein_2007,Bloch_2008}.
Moreover, disordered potentials can be engineered almost at will,
and their statistical properties can be controlled and tuned \cite{Clement_2006}.
So far, a large body of theoretical work has been devoted to
Anderson localization of non-interacting particles
\cite{Damski_2003,Roth_2003,Gavish_2005,Kuhn_2005,Massignan_2006,Sanchez-Palencia_2007,Kuhn_2007,Skipetrov_2008,Miniatura_2009,Pezze_2009,Pezze_2010,Antezza_2010,Piraud_2011}, interacting Bose \cite{Bilas_2006,Lugan_2007a,Lugan_2007b,Lugan_2011,Paul_2007,Horstmann_2007,Roux_2008,Paul_2009,Orso_2009,Falco_2009,Pollet_2009,Aleiner_2009} and Fermi \cite{Orso_2007,Byczuk_2009,Byczuk_2010,Han_2009}
disordered gases, and two-component disordered systems \cite{Sanpera_2004,Ahufinger_2005,Wehr_2006,Niederberger_2008,Niederberger_2009,Niederberger_2010,Crepin_2010}.
Experimentally, classical suppression of transport of Bose-Einstein condensates \cite{Clement_2005,Fort_2005}
and Anderson localization of matter-waves \cite{Billy_2008, Roati_2008} in one dimension (1D)
have been reported recently.
The competition between interaction and disorder has been investigated
in 1D bi-chromatic optical lattices \cite{Deissler_2009, Lucioni_2011} and
in three dimensional (3D) disordered lattices \cite{Pasienski_2009, White_2009}.

Studying the dynamics of even non-interacting atoms in a
disordered potential is a difficult problem, especially in dimensions higher than one.
For classical particles, the motion in disordered potentials is
generally chaotic, which sparks a variety of dynamical regimes \cite{Bouchaud_1990},
including classical localization, normal diffusion as well as sub- and super-diffusion.
For waves, diffusion is obtained in the limit of incoherent multiple 
scatterings \cite{VanRossum_1999,Vollhardt_1981,Shapiro_2007,Beilin_2010}.
Multiple coherent scatterings \cite{VanRossum_1999,Rammer_book,Kramer_1993} from the random defects 
can induce weak \cite{Hartung_2008} or strong \cite{Kuhn_2007} localization effects.
With a view towards search for localization effects in cold-atomic gases,
the characterization of classical transport regimes is a central task.
It is of special interest in dimension two (2D), which is the marginal
dimension for return probability in Brownian motion and for Anderson localization \cite{Abrahams_1979}.
In the context of cold-atomic gases, diffusion has been experimentally studied in random
dissipative 3D optical molasses \cite{Horak_1998,Grynberg_2000},
and conservative 2D random potentials \cite{Robert_2010}.

In this work, we numerically study the transport properties
of cold atoms in 2D disorder.
We consider a speckle potential and focus on the classical regime such that
\be \label{clreg}
\lambda_\textrm{\tiny dB} \ll \sigmar \leq l_\textrm{\tiny B} \ll L \ll L_{\mathrm{loc}},
\ee
where $\sigmar$ is the characteristic length scale of the disordered potential,
$\lambda_\textrm{\tiny dB}$ is the atomic de Broglie wavelength, 
$l_\textrm{\tiny B}$ is the Boltzmann (transport) mean free path,
$L$ is the system size and $L_{\mathrm{loc}}$ is the localization length.
The inequality $\lambda_\textrm{\tiny dB} \ll l_\textrm{\tiny B}$
guarantees that wave interference effects, which are important in 2D, 
occur on a very large scale which will be assumed here to be much
larger than the system size and thus disregarded.
In particular, besides that, in 2D, all quantum states are localized \cite{Abrahams_1979},
the inequality $L \ll L_{\mathrm{loc}}$ allows us to neglect strong localization effects.
When neglecting wave interference effects, the main dynamical regime is 
provided by diffusion \cite{Akkermans_book, Vollhardt_1981}. 
For $\sigmar \leq  l_\textrm{\tiny B}$ (which is most often true),
the condition $\sigmar \lesssim \lambda_\textrm{\tiny dB}$ may be fulfilled.
In this case, the wave nature of the particles governs the scattering from each defect.
Here instead we consider the opposite condition $\lambda_\textrm{\tiny dB} \ll \sigmar$
meaning that the atoms scatter in the
disordered potential as purely classical particles,
hence satisfying the Newton equations of motion.
Finally, since $l_\textrm{\tiny B}$ is the typical 
length of ballistic trajectories, the inequality $l_\textrm{\tiny B} \ll L$
ensures that the number of scattering events in the system is sufficiently 
large to strongly affect the motion of particles.
As discussed in our paper, the regime indicated by Eq.~(\ref{clreg}) 
is relevant to some recent cold-atom experiments \cite{Robert_2010}.

The central aim of our paper is to analyze the transport regimes
of classical particles at different energy.
In particular, we show that they strongly depend on
the topographic properties of the disordered potential.
An additional aim is to use these results to determine the behavior of
atomic clouds with a broad energy distribution, as relevant for
cold-atom experiments.

\begin{figure*}[t]
\begin{center}
\includegraphics[scale=0.24]{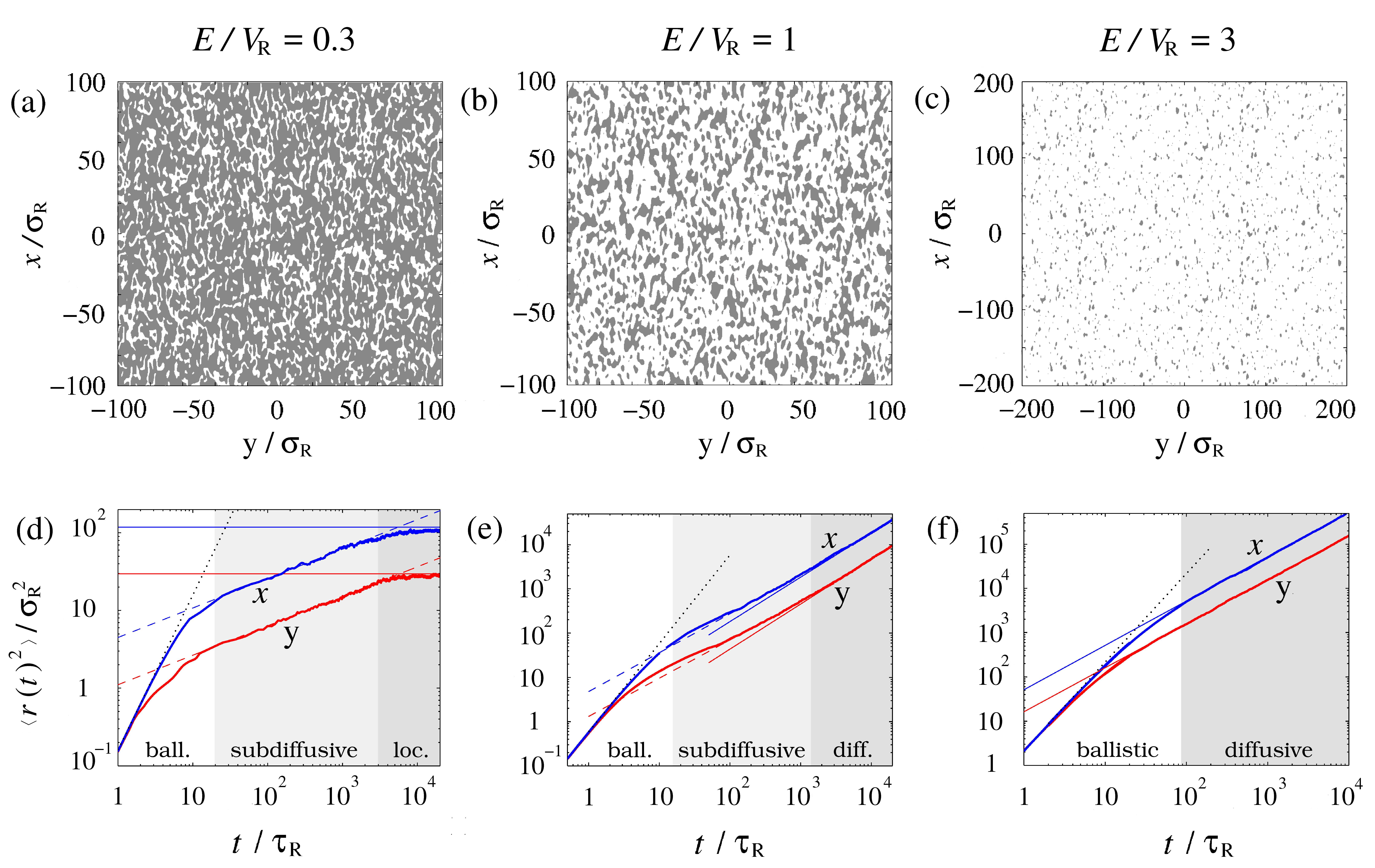}
\end{center}
\caption{\small{(color online) 
Topography and transport regimes
of classical particles in a 2D, anisotropic,
blue-detuned speckle potential with the anisotropy factor
$\lambda = \sigma_x/\sigma_y = 2$ (see Sec.~\ref{SpeckleStat}).
Panels (a)-(c) show the
allowed (white; $V(\vecr) \leq E$) and
forbidden (black; $V(\vecr) > E$) regions for
a classical particle of energy $E$ in the disordered potential $V(\vecr)$.
Different columns correspond to the same realization of the disordered potential
and different energies:
(a) and (d) $E=0.3\Vr$; (b) and (e) $E=1\Vr$; (c) and (f) $E=3\Vr$, 
where $\Vr$ is the average amplitude of the disordered potential
(see Sec.~\ref{SpeckleStat}).
In panel (a), the energy is below the percolation threshold
$E_\textrm{c}=0.52\Vr$ (for the 2D blue-detuned speckle potential used here, see Sec.~\ref{topology})
and all the particles are classically localized.
In panels (b) and (c), the energy is above the percolation threshold
and there appear allowed regions connecting the opposite sides of the box.
Panels (d)-(f) show the mean-square displacement,
$\big\langle r(t)^2 \big\rangle$
(the direction $r=x$ is represented by the upper thick blue line, $r=y$ by the lower thick red line)
as a function of time in units of $\taur \equiv \sqrt{m \sigmar^2/\vert \Vr \vert}$, 
where $m$ is the mass of the atoms (see Sec.~\ref{EquationMotion}).
The black dotted line is the isotropic ballistic behavior,
$\langle r(t)^2 \rangle \propto t^2$ predicted by Eq.~(\ref{ballistic}).
The dashed lines are
$\langle r(t)^2 \rangle \propto t^{\alpha_r}$
[see Eq.~(\ref{rt})] for intermediate time scales giving
a sub-diffusive behavior ($\alpha_r <1$) in (d) and (e).
The solid lines are $\langle r(t)^2 \rangle \propto t^{\alpha_r}$
for long times giving a time independent behavior ($\alpha_r =0$) in (d) and a
diffusive scaling ($\alpha_r=1$) in (e) and (f).}}
\label{Figure:Results}
\end{figure*}

\subsection{Main results presented in the paper.}
Here, we give a general overview of the main results
of this manuscript (leaving details for the text).
Figure~\ref{Figure:Results} illustrates the results
obtained for an anisotropic speckle potential
of anisotropy factor $\lambda = \sigma_x/\sigma_y$,
with $\sigma_r$ the correlation length in the $r\in\{x,y\}$ direction.
At any energy, a particle first undergoes a ballistic flight
for a typical time $\tau^* \sim l_\textrm{B}/v$ where $v$ is its initial velocity.
At sufficiently small energy, $E$, all classically allowed regions
in the disordered potential have a finite size [see Fig.~\ref{Figure:Results}(a)],
and the particle remains classically localized in the one containing its initial position.
For a particle initially at the origin, the disorder-averaged mean square displacement,
$\langle r(t)^2 \rangle \propto t^{\alpha_r}$,
shows a sub-diffusive behavior ($\alpha_r<1$) at intermediate times,
which corresponds to the random exploration by the particle of
its classically allowed region.
Once the full region is explored, $\langle r(t)^2 \rangle$ 
saturates asymptotically in time ($\alpha_r=0$) to a finite value
[see Fig.~\ref{Figure:Results}(d)].
Being of purely topographical nature, this behavior leads
to $\langle x^2 \rangle / \langle y^2 \rangle = \lambda^2$.
A critical energy $E_\textrm{c}$ signals the percolation phase transition corresponding to 
the appearance of infinitely extended allowed regions.
{For $E$ slightly above $E_\textrm{c}$, the percolating cluster
(classically-allowed region) is characterized by large lakes
connected by narrow bottlenecks [see Fig.~\ref{Figure:Results}(b)],
which are located at the saddle points of the disordered potential.
The particle then spends long times exploring a lake, with a similar
dynamics as described above, leading to the sub-diffusive behavior
observed in Fig.~\ref{Figure:Results}(e)
at intermediate (but experimentally relevant) times.
For longer times and larger distances, the dynamics is dominated
by rare but possible transitions between the lakes,
which leads, asymptotically in time, to normal diffusion,
$\langle r(t)^2 \rangle = 2 D_r t$.}
At larger energy, the lakes merge and the size of forbidden regions rapidly decreases 
[see Fig.~\ref{Figure:Results}(c)].
Then, the dynamics is characterized by normal diffusion
for $t>\tau^*$ [see Fig.~\ref{Figure:Results}(f)].
In this regime, we identify a power-law behavior of the diffusion 
coefficients as a function of the particle energy
and show that the anisotropy of diffusion significantly differs from
the anisotropy of the disorder, $D_x/D_y \neq \lambda^2$.

It is worth noting that a sub-diffusion regime
should not be viewed as a precursor of localization.
In fact, we see in the example considered here
that sub-diffusion can cross over to
either localization (for $E \lesssim E_\textrm{c}$)
or normal diffusion (for $E \gtrsim E_\textrm{c}$).

\subsection{Outlook.}
The manuscript is organized as follows.
In Sec.~\ref{DisorderedPotential},
we acquaint with the main statistical and topographic properties 
of the 2D speckle potential used along the paper.
We identify a percolation phase transition and determine 
the critical exponent characterizing the divergence of the classical localization length.
In Sec.~\ref{FixedEnergy}, we analyze the dynamics 
of non-interacting classical particles of given energy 
in the disordered potential.
Firstly, we write down the classical 
equations of motion and, with a proper rescaling, 
we highlight the universality class of the classical dynamics.
Due to the redistribution of kinetic and potential energy 
induced by the disordered potential,
the total particle energy is identified as the sole relevant parameter
for the dynamics.
Different transport regimes are then identified and discussed:
(i)~classical localization, 
(ii)~normal diffusion and
(iii)~transient subdiffusion.
Secondly, in Sec.~\ref{AtomicCloud},
we use the above results as a groundwork to study the overall dynamics
of atomic gases with a broad energy distribution, as relevant to cold-atom experiments.
In particular, we show that, in the diffusion regime,
the mean square size of the gas grows linearly in time
($\langle r(t)^2 \rangle \propto t$)
but, due to the summation of all diffusive components
with energy-dependent diffusion coefficients,
expanding density profiles significantly deviate from Gaussian functions.
In Sec.~\ref{Exp}, considering a thermal cloud released in the disordered potential,
we compare our predictions with the experimental results of Ref.~\cite{Robert_2010}.
We provide fitting functions to extract the single-energy diffusion coefficients
from the collective expansion of an atomic cloud of atoms. 
We find fair agreement between numerical calculations and experimental measurements.
Finally, we summarize our findings and discuss possible extensions of
this work in Sec.~\ref{Conclusions}.
On the one hand, our analysis provides a guideline for the experimental investigation
of classical transport in cold atomic gas in disordered potentials~\cite{Robert_2010}.
On the other hand, it may contribute to the interpretation of on-going
experiments searching for deviation from classical diffusion and traces of
Anderson localization with ultracold gases in 2D geometry.

\section{Disordered Potential}
\label{DisorderedPotential}

\subsection{Statistical properties of the speckle potential}
\label{SpeckleStat}

Throughout the paper, we consider a speckle disorder,
as commonly devised in cold-atom experiments
\cite{Fort_2005, Clement_2005, Clement_2006, Billy_2008,
Clement_2008, Chen_2008, Dries_2010, Robert_2010}.
A speckle pattern is created, for instance, by
a coherent monochromatic laser beam passing through
a diffusive plate \cite{Clement_2006}.
A fully developed speckle field can be produced by a ground glass
whose surface contains random grains associated to optical path length
fluctuations longer than the laser wavelength.
In this case, the speckle field, observed
at the focal plane of a converging lens, results from the
coherent superposition of waves originating from different
scattering sites on the rough surface with uncorrelated
phases uniformly distributed in $[0, 2\pi]$.
Owing to the central limit theorem, the real and imaginary parts of the
electric field are independent Gaussian variables.
The light-shift potential $V(\vecr)$ felt by the atoms exposed to the electromagnetic
radiation is proportional to the field intensity:
it is thus a random potential but we stress that its statistics is not Gaussian.
The speckle potential is also inversely proportional to the detuning of the laser light
with respect to the two-level atomic transition
and can be either attractive or repulsive (see below).
Finally, assuming a sufficiently far-detuned laser field, we will neglect,
in the following, dissipative effects.

\begin{figure}[!t]
\begin{center}
\includegraphics[scale=0.3]{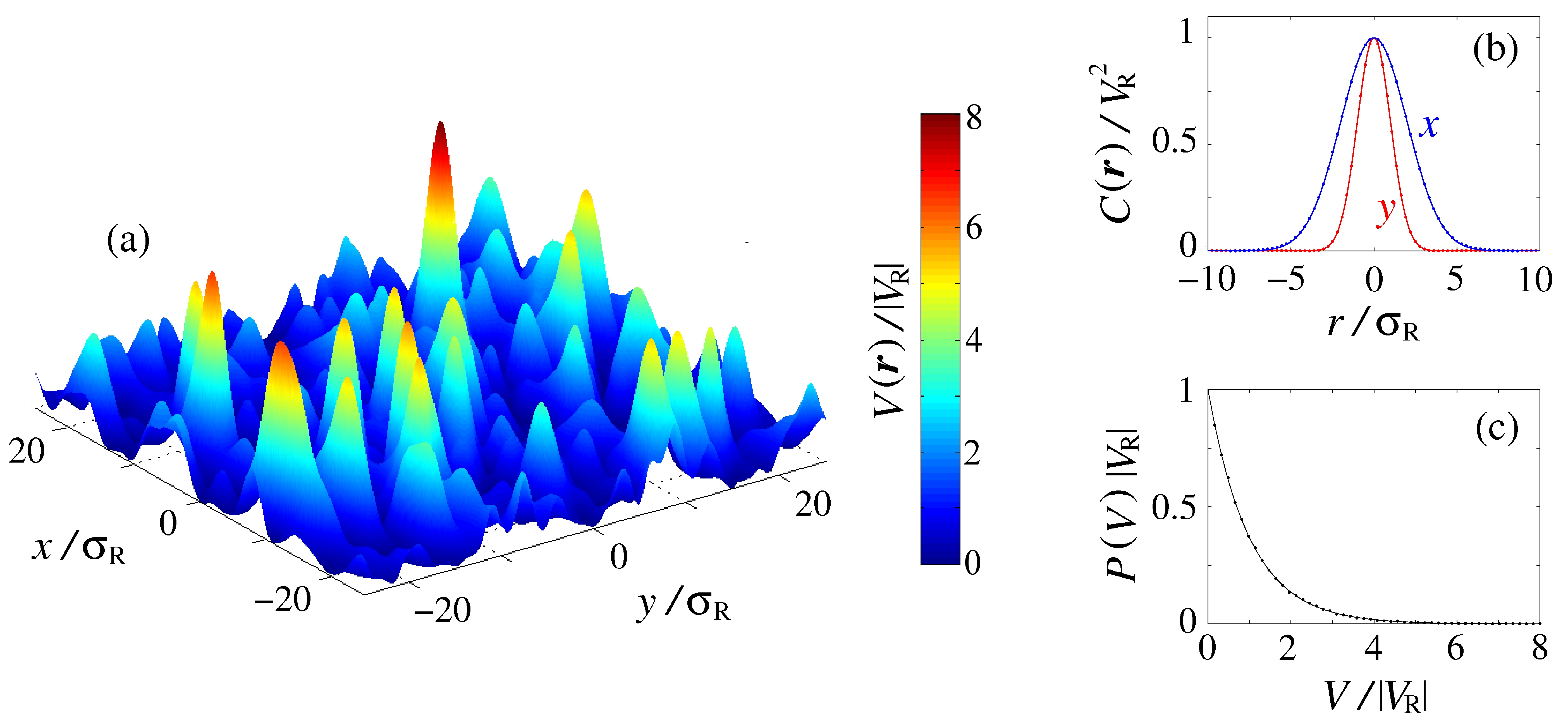}
\end{center}
\caption{
\small{(color online) (a) Example of 
anisotropic speckle potential used in the numerical simulations.
The anisotropy factor is $\lambda=\sigma_x/\sigma_y=2$.
(b) Autocorrelation function of the disordered potential (dots), 
along the $x$ and $y$ directions [$C(x,y=0)$ and $C(x=0,y)$, respectively]. 
It has been obtained numerically by averaging over 100000 realizations of the disordered
potential. The solid lines correspond to Eq.~(\ref{Corrf}) with no fitting parameters.
(c) Intensity distribution obtained for a single realization of the disordered 
speckle potential (dots). The solid line is Eq.~(\ref{PV}).}}
\label{Figure:Speckle}
\end{figure}

The statistical properties of the speckle potential have been extensively
investigated \cite{Goodman_book}.
The probability that the potential, at a given point $\vecr$,
has amplitude $V \equiv V(\vecr)$ follows an exponential law
\be \label{PV}
P(V) = \frac{e^{-V/\Vr}}{\vert \Vr \vert} \, \Theta\bigg( \frac{V}{\Vr} \bigg),
\ee
where $\Theta$ is the Heaviside step function.
We have $\langle V(\vecr) \rangle=\Vr$ and
$(\Delta V(\vecr))^2 \equiv \langle V(\vecr)^2 \rangle-\langle V(\vecr) \rangle^2=\Vr^2$,
where $\langle ... \rangle$ indicates the average over the different realizations of the
disordered potential.
The quantity $\Vr$ is the amplitude of the disorder,
and its sign depends on the detuning of the laser light from the atomic resonance:
For a blue-detuned speckle potential,
the disorder is repulsive (i.e.\ positive $\Vr > 0$ and $V(\vecr) \geq 0$)
and it is characterized by arbitrary large intensity peaks with exponentially low probability.
Conversely, for a red-detuned speckle potential,
the disorder is attractive (i.e.\ negative $\Vr < 0$ and $V(\vecr) \leq 0$)
and it is characterized by arbitrary low dips with exponentially low probability.

The autocorrelation function of the disordered potential,
$C(\vecr) = \langle V(\vecr+\vecr') V(\vecr') \rangle - \langle V(\vecr') \rangle^2$,
can be controlled by the geometry of the diffusive plate \cite{Clement_2006}.
Without loss of generality, it can be written as
$C(\vecr) = \Vr^2 c(\vecr/\sigmar)$ \cite{note_corrf},
where $\sigmar$ is the correlation length of
the disordered potential, that is
the typical width of $C(\vecr)$,
which is proportional to the speckle grain size.
The precise definition of $\sigmar$ depends on the specific model of disorder. 
In this manuscript, we will focus on a 2D, anisotropic, speckle potential 
with a Gaussian correlation function of widths $\sigma_x$ and $\sigma_y$ 
along the two main directions,
\be \label{Corrf}
C(\vecr) = \Vr^2 \, \exp\big(-x^2/2\sigma_x^2\big) \exp\big(-y^2/2\sigma_y^2\big).
\ee
We define $\sigmar \equiv \sigma_x$ and
$\lambda \equiv \sigma_x/\sigma_y$ the anisotropy factor.
In Fig.~\ref{Figure:Speckle}(a), we show a typical realization of
a speckle potential as used in the numerical simulations ($\lambda=2$).
Figures \ref{Figure:Speckle}(b) and \ref{Figure:Speckle}(c)
show its autocorrelation function and intensity distribution, respectively.
Experimentally, a speckle potential with a Gaussian correlation function 
can be obtained by illuminating a diffusive plate with a Gaussian laser beam.
The anisotropy can be controlled by the aperture function of the diffusive 
plate~\cite{Goodman_book, Clement_2006}
or by tilting a 2D, isotropic speckle field with respect to the
diffusion plane of the atoms~\cite{Robert_2010}.

\subsection{Topographic properties of the speckle potential}
\label{topology}
We now study some topographic properties of the 2D speckle potential
which are relevant to understand the transport regimes of classical particles
[see Sec.~\ref{FixedEnergy}].
Some of them, as, for instance,
the percolation threshold, which will be discussed below,
do not depend on the anisotropy of the potential.
Therefore, these are straightforwardly obtained from those of
the isotropic case, upon rescaling of the $y$ direction by the anisotropy factor $\lambda$.
We thus focus, without loss of generality, on an isotropic speckle disorder
in this sub-section.

The positivity of the kinetic energy constraints the trajectories of
classical particles {with total energy $E$}
to the sub-space defined by $V(\vecr) \leq E$.
This condition separates the space in allowed and forbidden regions.
For the sake of illustration, Figs.~\ref{Figure:Results}(a)-(c)
show the classically-allowed ($V(\vecr) \leq E$; white) and
forbidden ($V(\vecr) > E$; black) regions for particles of fixed 
energy in a specific realization of a 2D blue-detuned anisotropic speckle potential.
In a continuous disordered potential, the topography of classically allowed regions
strongly depends on the ratio $E/\vert\Vr\vert$, 
and, for a speckle potential, on the sign of $\Vr$.
The fraction of space satisfying the condition $V(\vecr) \leq E$
is given by~\cite{Zallen_1971}
\be \label{AreaFraction}
\phi(E) = \int_{-\infty}^{E} \ud V \, P(V).
\ee
For a speckle potential, $P(V)$ is given by Eq.~(\ref{PV}).
For blue-detuned speckle potentials, 
we thus find $\phi(E) = 1 - e^{-E/\vert \Vr \vert}$
for $E \geq 0$ and $\phi(E)=0$ (i.e. there is no allowed region) for $E \leq 0$.
For red-detuned speckle potentials, we have 
$\phi(E) = e^{+E/\vert \Vr \vert}$ for $E\leq 0$
and $\phi(E)=1$ (i.e. there are no forbidden region) for $E\geq 0$.

For 2D disordered potentials, 
there exists a critical energy, the
so-called \textit{percolation threshold}, $E_\textrm{c}$,
which marks the sharp transition between the existence of infinitely extended
allowed and forbidden regions \cite{Zallen_1971}.
The \textit{percolation transition}
is often considered as the simplest possible phase transition
with nontrivial critical behavior~\cite{Isichenko_1992}.
It has a pure geometrical nature and
exhibits long-range correlation near the critical value, corresponding
to the divergence of the average size of the allowed regions (in infinite space).

Below $E_\textrm{c}$, the allowed regions are disconnected and 
form ``lakes'' of finite size [see Fig.~\ref{Figure:Results}(a)].
The allowed regions consist of a finite number of minima of the potential which are
connected through saddle points. 
In other words, particles of energy $E \leq E_\textrm{c}$ are classically localized,
i.e. they cannot move to arbitrary large distances.
On a macroscopic scale, the system behaves as an insulator.
Above $E_\textrm{c}$, an infinite number of ``lakes'' merge together to 
form an infinite ``ocean'': 
The allowed subspace contains at least
one connected subset (cluster) which stretches to infinity
[see Figs.~\ref{Figure:Results}(b)-(c)].
In other words, particles with energy $E > E_\textrm{c}$ and 
initially placed in an infinitely-extended allowed region 
can move to infinity. 
The system thus behaves as a conductor.

\begin{figure}[!t]
\begin{center}
\includegraphics[scale=0.58]{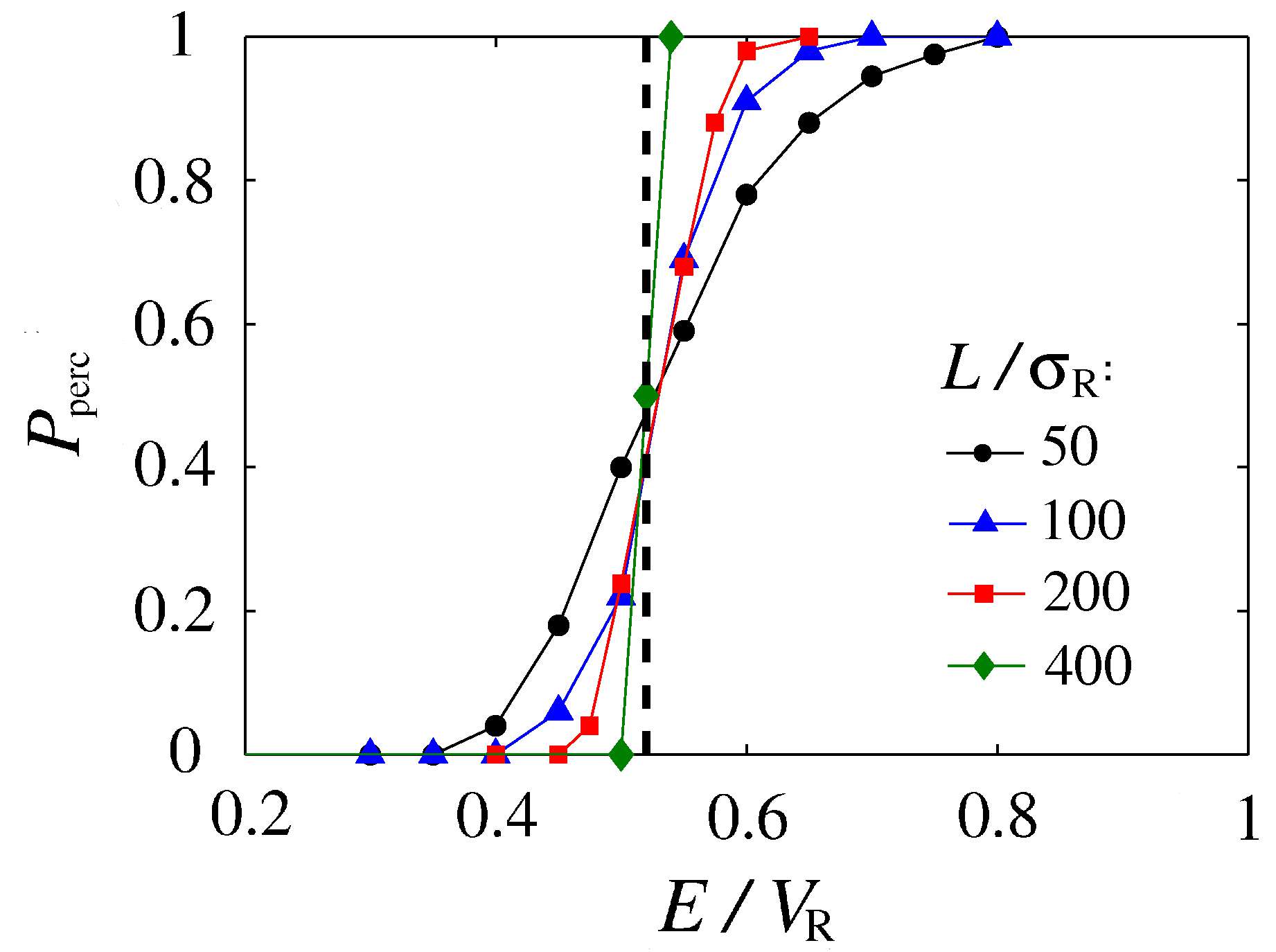}
\end{center}
\caption{\small{(color online) Percolation probability 
as a function of the particle energy.
It is calculated numerically as the probability to find 
at least one allowed region connecting the opposite sides of  
a square box of size $L$ in a blue-detuned (isotropic) speckle potential.
Different symbols correspond to different 
values of $L$ (indicated in the figure)
and the solid lines are guides to the eye.
The vertical dashed line indicates the percolation threshold, $E_\textrm{c}=0.52\Vr$.
The percolation probability is calculated over 100 realizations of the disorder 
and the numerical grid step is $\sigma_R/10$.}}
\label{Figure:PercProb}
\end{figure}

A challenging issue is to determine the value of the
percolation threshold $E_\textrm{c}$ for different models of disorder.
In the case of a disordered potential with sign symmetry,
[i.e. when the statistical properties of the
potential $V(\vecr)$ are equivalent to those of $2 \langle V(\vecr) \rangle - V(\vecr)$],
the percolation threshold is $E_\textrm{c} = \langle V(\vecr) \rangle$ and 
the corresponding critical area fraction is
$\phi_\textrm{c} \equiv \phi(E_\textrm{c})=1/2$~\cite{Zallen_1971}.
An example is given by Gaussian random distributions.
However, a speckle potential is non-Gaussian and
does not have the sign symmetry discussed above.
For instance, a blue-detuned speckle potential is bounded below by zero,
but it is unbounded above (and {\it vice versa} for the red-detuned speckle
potential).
We have numerically investigated the percolation threshold in 2D speckle potentials
with the correlation function given by Eq.~(\ref{Corrf})
{with $\lambda=1$}.
In particular, we have evaluated the probability 
(for different realizations of the disorder),
$P_{\mathrm{perc}}$, to have an allowed region connecting the opposite sides of a square box.
The results are shown in Fig. \ref{Figure:PercProb}
for a blue-detuned speckle and for different box lengths, $L$.
As expected, the crossover between $P_{\mathrm{perc}}=0$ and $P_{\mathrm{perc}}=1$
becomes sharper and sharper by increasing $L$,
and all curves cross the value $P_{\mathrm{perc}}=0.5$ at approximately the same
value of $E/\Vr$. This is compatible with a phase transition at $E_\textrm{c}=0.52\Vr$,
for a blue-detuned, 2D, speckle potential,
in the limit $L \rightarrow \infty$ (vertical dashed line in the figure).
According to Eq.~(\ref{AreaFraction}), the critical fraction of allowed regions
is given by $\phi_\textrm{c}=0.405$, a value significantly different
from $0.5$ as expected for potentials with sign-symmetry.
The values $E_\textrm{c}=0.52\Vr$ and $\phi_\textrm{c}=0.405$ found here 
are in agreement with the experimental findings~\cite{Smith_1979}
and numerical results \cite{Weinrib_1982} obtained by
mapping the speckle potential onto a network
connecting potential minima through saddle points.
Because of the duality between allowed and forbidden regions in the 2D case,
when inverting the sign of the potential, allowed regions for
a particle of energy $E$ becomes forbidden regions for a
particle of energy $-E$.
The percolation threshold for the red-detuned speckle potential can then
be directly calculated from that of the blue-detuned speckle potential: $E_\textrm{c}(-\Vr)= - E_\textrm{c}(\Vr)$.
For red-detuned speckle potentials, we thus find that the percolation 
threshold is at $E_\textrm{c}=-0.52\vert\Vr\vert$
and the corresponding fraction of allowed regions is $\phi_\textrm{c}=0.595$.

\begin{figure}[!t]
\begin{center}
\includegraphics[scale=0.35]{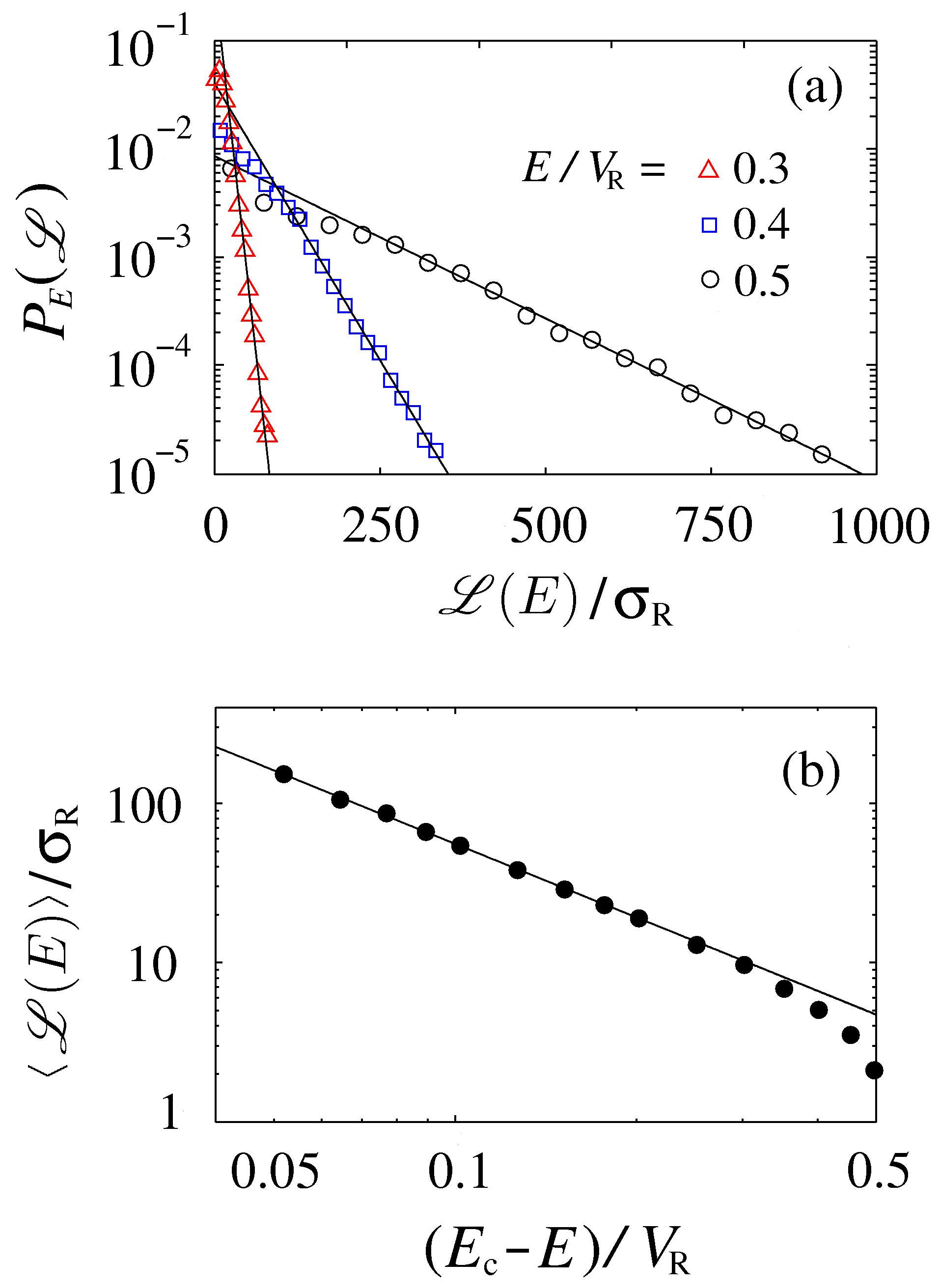}
\end{center}
\caption{\small{(color online)
(a) Probability distribution of the classical localization 
length defined in Eq.~(\ref{CLoc}), for a blue-detuned speckle potential.
Symbols are numerical results for different energies (indicated in the figure).
The solid lines are fits to an exponential function.
For instance, for $E=0.3\Vr$ (triangles), the fit gives 
$P_E\big(\mathscr{L}\big) \simeq (0.35/\sigmar) \, \exp(-0.12 \, \mathscr{L}/\sigma_R)$
for $\mathscr{L} \gtrsim 30 \sigma_R$.
(b) Average localization length as a function of $E_\textrm{c}-E$. 
{The line is a fit} to Eq.~(\ref{criticaL})
where $a$, $E_\textrm{c}$ and $\nu_c$ are the fitting parameters.
The numerical results (dots) have been obtained after averaging over 10000 
realizations of a blue-detuned isotropic speckle 
potential in a box of length $L=800 \sigma_R$ with grid step $\sigma_R/8$.}}
\label{Figure:Loc}
\end{figure}

In order to further characterize the topographic properties
of the speckle potential in a homogeneous box 
we have studied the classical localization 
length $\mathscr{L}(E)$, i.e. the mean diameter of the percolating cluster
stretching from the box center.
We use the definition
{
\be \label{CLoc}
\mathscr{L}(E) \equiv 2\sqrt{ \delta\vecr^2},
\ee
where $\delta\vecr^2 = \int \ud^2\vecr\, A_E(\vecr) \vert\vecr\vert^2
/ \int \ud^2 \vecr\, A_E(\vecr)$
with $A_E(\vecr)=1$ in the
classically allowed region stretching from the box center [if $V(\vecr=0)\leq E$]
and $A_E(\vecr)=0$ elsewhere.}
The calculation of the localization length, Eq.~(\ref{CLoc}), 
at fixed energy is repeated for several realizations of the disordered potential.
The corresponding probability distribution $P_E\big(\mathscr{L}\big)$
is shown in Fig.~\ref{Figure:Loc}(a), for various energies.
It decreases exponentially for sufficiently large values of $\mathscr{L}$.
The average localization length, 
$\langle \mathscr{L}(E) \rangle \equiv \int \ud \mathscr{L} \, \mathscr{L} P_E\big(\mathscr{L}\big)$ 
is shown in Fig.~\ref{Figure:Loc}(b).
It diverges when approaching the percolation threshold $E_\textrm{c}$ and, 
close to this value, it can be well fitted to the function
\be \label{criticaL}
\langle \mathscr{L}(E) \rangle = a \bigg(\frac{E_\textrm{c} -E}{\Vr}\bigg)^{-\nu_c},
\ee
where $a=1.63 \pm 0.3 \, \sigma_R$, the critical exponent 
$\nu_c = 1.53 \pm 0.2$ and the percolation threshold $E_\textrm{c}=0.54 \pm 0.02$
are fitting parameters.
The value of $E_\textrm{c}$ obtained from the fit is 
in agreement, within the numerical error, with the value $E_\textrm{c}=0.52 \Vr$ 
found with other methods [see Fig.~\ref{Figure:PercProb}].

\begin{figure}[!t]
\begin{center}
\includegraphics[scale=0.37]{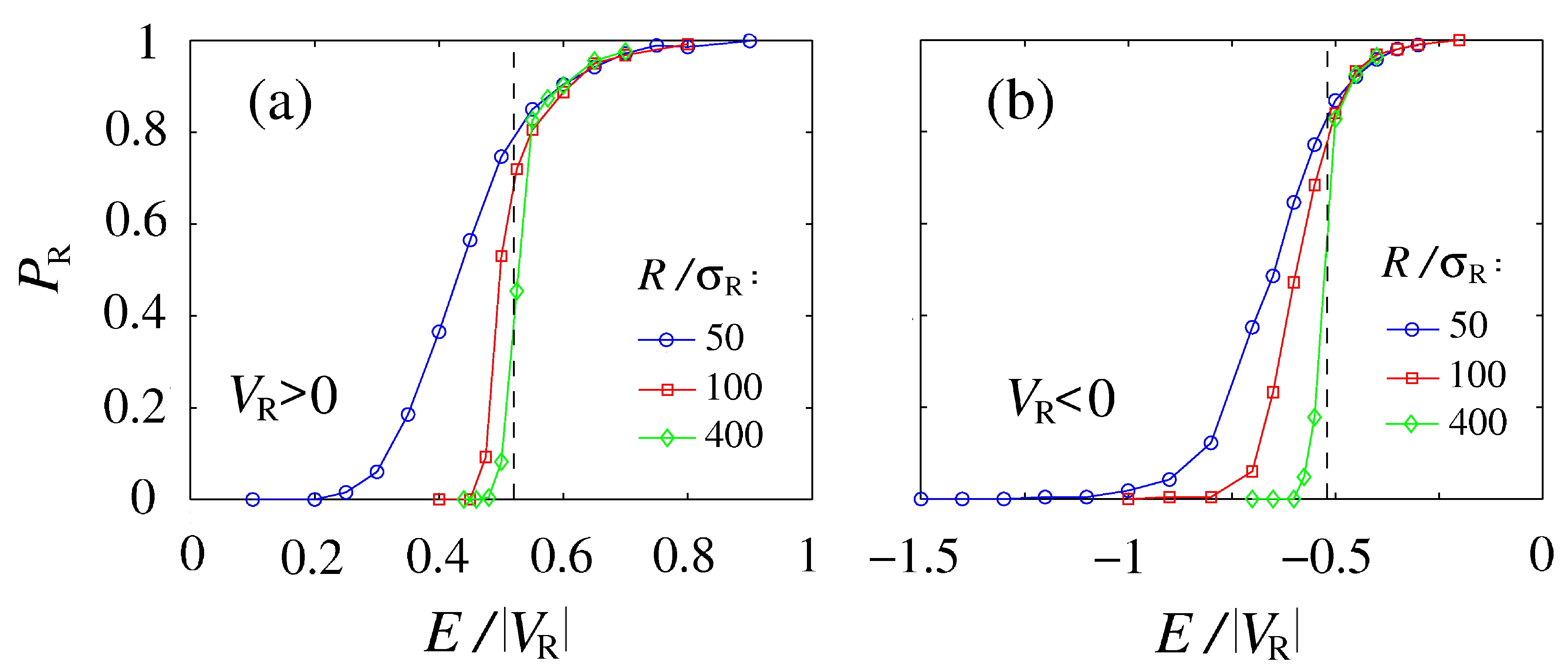}
\end{center}
\caption{
\small{(color online) Probability to find a classically-allowed region
stretching to a distance larger than $R$.
Each point has been obtained after averaging over 500 realizations of
the disordered potential.
Different lines refer to different values of $R$ (indicated in the figure).
Panels (a) and (b) correspond
to blue- and red-detuned speckle potentials, respectively.
The vertical dashed lines are the percolation threshold $E_\textrm{c}=0.52 \Vr$.}}
\label{Figure:PercLoc}
\end{figure}

Above the percolation threshold, for $E>E_\textrm{c}$,
not all the allowed energy regions extend to infinity 
and some particles can still be classically localized, 
depending on their initial position.
This behavior is relevant to our system, where the 
dynamics starts {\it inside} the disordered potential.
We have thus studied, as a function of the particle energy,
the probability $P_R$ to find, for different realizations of the disordered potential, 
an allowed region connecting two points at a distance $R=\sqrt{x^2+y^2 }$.
The results are shown in Figs.~\ref{Figure:PercLoc}(a) and (b) for the blue- and 
red-detuned speckle potentials, respectively.
For large enough distance $R$, the probability shows a sharp transition,
from zero, for $E \leq E_\textrm{c}$, to a finite value, for $E > E_\textrm{c}$.
Note that, for $E > E_\textrm{c}$, the probability converges to a
curve which does not depend on the length $R$.
This is consistent with the existence of 
allowed regions of finite size above the percolation threshold.
The number of such regions rapidly decreases when increasing the energy. 
In a blue-detuned speckle potential [see Fig.~\ref{Figure:PercLoc}(a)],
there are always forbidden regions but, 
for energy $E \gtrsim \Vr$, almost all percolation regions extend to infinity 
($P_R$ quickly approaches the value 1).
In a red-detuned speckle potential [see Fig.~\ref{Figure:PercLoc}(b)]
there is a single percolation region for $E>0$, as there is no forbidden region in this case.

\section{Transport regimes in a 2D disordered potential}
\label{FixedEnergy}

In this section, we study the dynamics of a classical particle 
of fixed total energy $E$ in a 2D anisotropic disorder.
The central goal is to characterize the behavior of the mean square displacement 
as a function of the particle energy. 
We mainly focus on blue-detuned speckle potentials
but also discuss results for the red-detuned case.

\subsection{Classical equations of motion}
\label{EquationMotion}

The dynamics of a classical particle in
a disordered potential is most conveniently written
by rescaling the dynamical variables in the following way:
\begin{subequations} \label{rescaling}
  \begin{gather}
    \tilde{E}=E/\vert\Vr\vert,       \label{rescaling.1} \\
    \tilde{\vecr} =  \vecr/\sigmar,         \label{rescaling.2} \\
    \tilde{\vecp} =  \vecp/\pr,         \label{rescaling.3} \\
    \tilde{t}     =  t /\taur,      \label{rescaling.4}
  \end{gather}
\end{subequations}
where $E$ is the particle energy, $\vecr=(x,y)$ its position,
$\vecp=(p_x,p_y)$ its momentum, and $t$ is the time.
Introducing the momentum
$\pr = \sqrt{m \vert\Vr\vert}$
and the time
$\taur = \sqrt{m \sigmar^2/\vert\Vr\vert}$, where
$m$ is the atomic mass, the classical equations of
motion read:
\begin{subequations} \label{ClEq}
  \begin{gather}
    \frac{\ud \tilde{\vecr}}{\ud \tilde{t}} = \tilde{\vecp},                                                            \label{ClEq.1} \\
    \frac{\ud \tilde{\vecp}}{\ud \tilde{t}} = - \frac{\ud \upsilon(\tilde{\vecr})}{\ud \tilde{\vecr}}, \label{ClEq.2}
  \end{gather}
\end{subequations}
where $\upsilon (\tilde{\vecr})=V(\vecr)/\vert\Vr\vert$ is the reduced disordered potential.
Equations (\ref{ClEq}) highlight the universality class of the classical dynamics:
upon the above rescaling by the disorder parameters $\Vr$ and $\sigmar$,
the classical dynamics only depends on the properties of the reduced potential,
$\upsilon(\tilde{\vecr})$, that is on the model of disorder,
{and in particular on its anisotropy}.
It is then sufficient to study the particle dynamics for a single set of
parameters $\Vr$ and $\sigma_R$.
This is an important difference compared to the
quantum dynamics, which also depends on
$\Vr/\Er$, where $\Er=\hbar^2/m\sigmar^2$ \cite{Kuhn_2007, Lugan_2009, Gurevich_2009}.

The reduced equations (\ref{ClEq}), for the speckle potential described in
Section \ref{DisorderedPotential}, form the basis of the calculations 
reported in this paper.
For each realization of the disordered potential $V(\vecr)$,
we numerically solve Eqs.~(\ref{ClEq}) to calculate 
the classical phase-space trajectory $(\vecr(t), \vecp(t))$,
depending on the initial conditions $(\vecr_0,\vecp_0)$.
We use an adaptive ordinary
differential equation algorithm where the time step is
adjusted such that the energy is conserved within about $0.5\%$
over the entire evolution.
Numerical simulations are run with the
fixed initial position $\vecr_0=0$.
The initial momentum is chosen with amplitude $|\vecp_0| = \sqrt{2m [E-V(\vecr_0) ]}$
and, unless specified, along a random direction chosen homogeneously.
Disordered speckle potentials are randomly generated and disregarded if $V(\vecr_0)>E$.
They are typically created in a box of linear length $L = 400 \sigmar$
with a grid step $\Delta x = \Delta y =\sigmar/10$ and repeated periodically over
the infinite plane \cite{nota_continuity_speckle}.
Quantities of interest (see below) are then obtained by
averaging over different realizations of the disordered potential.


\subsection{Time scale of ballistic dynamics}
\label{subsec_ballistic}

\begin{figure}[!t]
\begin{center}
\includegraphics[scale=0.17]{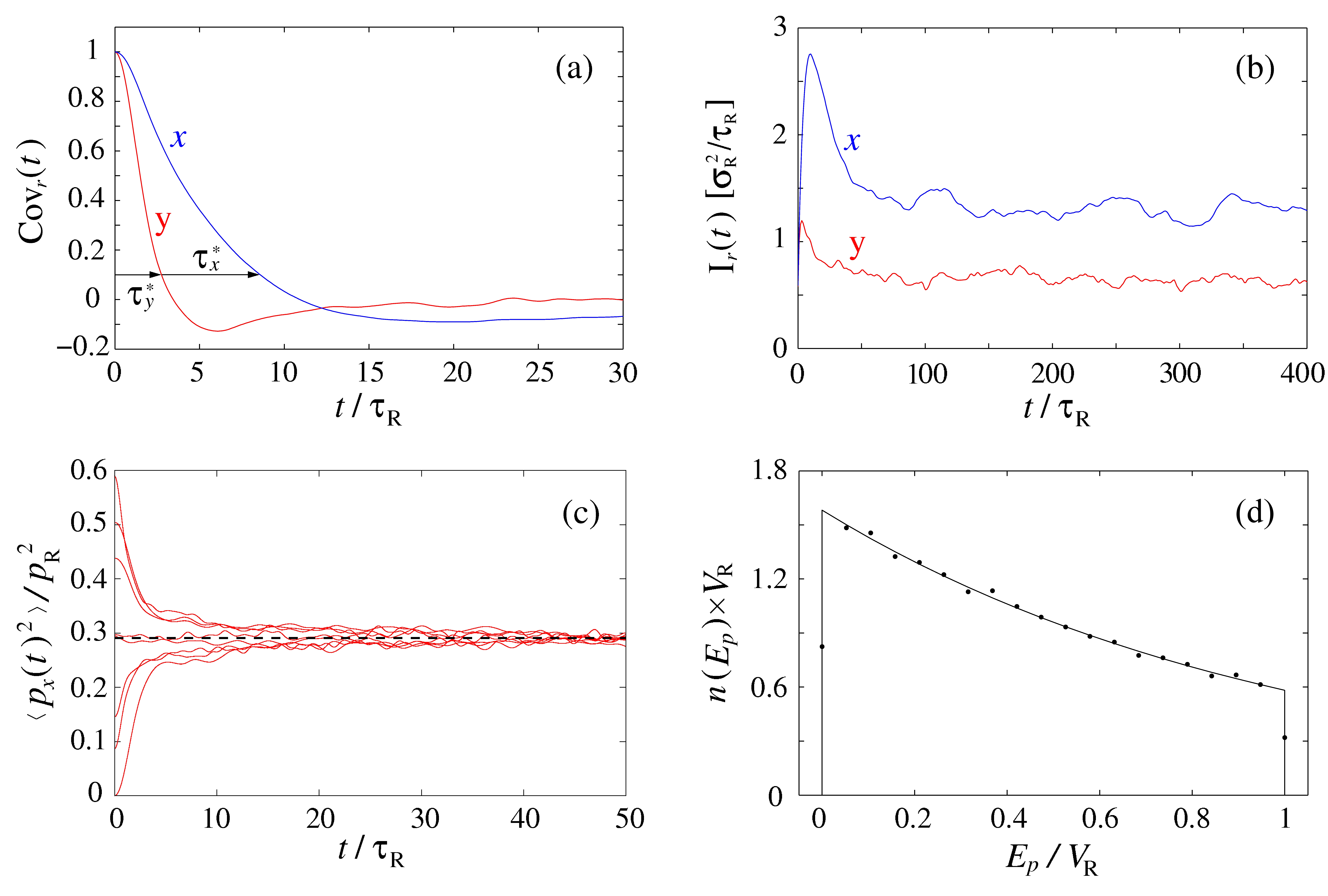}
\end{center}
\caption{\small{(color online)
Primary effects of the disordered potential on the dynamics of 
a classical particle: modification of the ballistic dynamics and 
redistribution of energy. Here we consider particles of energy $E=\Vr$.
Panel (a) shows the normalized momentum covariance $\mathrm{Cov}_{r}(t)$
for particles with momentum 
$p_r(0) = \sqrt{2m [E-V(\vecr_0) ]}$
along the $r=x$ (blue line) and $r=y$ (red line) direction, as a function of time.
(b) Integral of the momentum covariance,
$\mathrm{I}_r(t)$,
(in units $\sigmar^2/\taur$), as a function of time and 
along the $r=x$ (blue line) and $r=y$ (red line) direction.
(c) Average kinetic energy, $\langle p_x(t)^2 \rangle/2m$.
The different solid lines refer to particles with fixed energy and 
different initial momentum directions.
The dotted line is Eq.~(\ref{Ek1D_2D}).
Analogous results are obtained for $\langle p_y(t)^2 \rangle/2m$.
(d) Potential energy distribution obtained at time $t = 30 \taur$.
The dots are numerical results and the solid line is Eq.~(\ref{nV2D}).
The numerical results have been obtained for a blue-detuned speckle potential ($\Vr>0$)
by averaging over 10000 realizations in panels (a) and (d) and 2000 realizations
in panels (b) and (c).}}
\label{Figure:Edist}
\end{figure}

The primary effect of the disordered potential is to modify the 
ballistic dynamics of particles.
A particle of energy $E$ initially moving along the $r \in \{x,y\}$
direction looses the memory of its initial momentum
within a characteristic time scale $\tau_r^*(E)$ \cite{note_Boltzmann}.
In order to identify $\tau_r^*(E)$, we have studied the momentum covariance tensor 
\be \label{Cov}
\mathrm{Cov}_{r,r'}(t) \equiv \big\langle \, p_r(t) \, p_{r'}(0) \, \big\rangle. 
\ee
The typical behavior of the diagonal elements of the normalized 
momentum covariance,
$\mathrm{Cov}_{r}(t) \equiv \mathrm{Cov}_{r,r}(t) /\big \langle \, p^2_{r}(0) \, \big\rangle$, 
is shown in Fig.~\ref{Figure:Edist}(a) as a function of time
and along the $r=x,y$ directions.
We have checked that the off-diagonal elements of Eq.~(\ref{Cov})
vanish, in our case.
The quantity $\mathrm{Cov}_{r}(t)$ is characterized by a rapid decay and
we define $\tau_r^*(E)$ such that 
$\mathrm{Cov}_{r}\big[\tau_r^*(E)\big]=0.1$.
For an anisotropic speckle disorder, we find $\tau_x^*(E) \neq \tau_y^*(E)$ and
$\tau_r^*(E)$ increases with $\sigma_r$ ($r \in\{x,y\}$).
We then introduce $\tau^*(E) \equiv \mathrm{min}[\tau_x^*(E), \tau_y^*(E)]$.
For times $t \lesssim \tau^*(E)$, the dynamics is ballistic and isotropic.
For times $t \sim \tau^*(E)$, the dynamics is modified by the disorder,
it enters various transport regimes and may become anisotropic 
(see Sec.~\ref{DiffusionRegimes}).

Note that, in our simulations, $\mathrm{Cov}_{r}(t)$ 
does not exactly converge to zero at large times as exemplified in Fig.~\ref{Figure:Edist}(a).
We indeed find fluctuations with both positive and negative values, 
which survive massive configuration averaging.
We have however checked, from our numerical results, that the integral 
$\mathrm{I}_r(t) \equiv \int_{0}^{t}\ud u \, \mathrm{Cov}_{r,r}(u)/m^2$
converges in the long time limit to a finite value (up to fluctuations), 
as shown in Fig.~\ref{Figure:Edist}(b).
This holds for particles of energy both above and below the percolation threshold.
In addition, for particles in the normal diffusion regime, the limit of this 
integral equals the diffusion coefficient obtained with the fit of the 
mean-square displacement as discussed in Sec.~\ref{DiffusionRegimes}).
This is consistent with the celebrated relation,
$D_{r,r'}=\int_{0}^{+\infty} \, \ud t \,
\mathrm{Cov}_{r,r'}(t)/m^2 $,
valid for normal diffusion \cite{Isichenko_1992},
where $D_{r,r'}$ is the $(r,r')$ component of the diffusion tensor.

\subsection{Energy redistribution}
\label{redistribution}

Another main effect of disorder is to redistribute, 
after a time of the order of $\tau^*(E)$,
the different forms (kinetic and potential)
of the particle energy, under the constraint that 
the total energy is conserved.
An example is shown in Fig.~\ref{Figure:Edist}(c) where we 
plot the average kinetic energy along the $x$ direction, 
$\langle p_x(t)^2 \rangle/ 2m$, as a function of time. 
Different lines correspond to particles with the same energy $E=\Vr$
and different initial momentum directions $\vecp_0$.
After times of the order of $\tau^*(E)$, the lines 
converge to the same value, $\langle E_\textrm{k} \rangle$, 
which will be calculated below [see Eq.~(\ref{Ek1D_2D})].

A first insight into the dynamics is obtained
by using the statistical micro-canonical ensemble,
which assumes phase-space ergodicity in the shell of energy $E$.
The probability distribution of the potential energy $E_\textrm{p}$ 
for a particle of energy $E$ in a single realization of the disordered potential
is then 
$n(E_\textrm{p}) \propto
\int \ud \vecr \, \ud \vecp\, \delta[E-\vecp^2/2m - V(\vecr)] \, \delta[V(\vecr)-E_\textrm{p}]$.
In a continuous, homogeneous disorder, averaging over realizations of the disordered potential
equals spatial averaging, so that
$\int \ud \vecr\ \delta[V(\vecr)-V] \propto P(V)$ \cite{Lifshits_book}.

For a blue-detuned speckle potential, using Eq.~(\ref{PV}), we find,
\beq \label{nV2D}
n(E_\textrm{p}) =
\frac{e^{-E_\textrm{p}/\Vr}}{\Vr(1 - e^{-E/\Vr})} \Theta(E_\textrm{p}) \Theta(E - E_\textrm{p}),
~~\Vr>0,~~~~~
\eeq
normalized such that $\int \ud E_\textrm{p} \, n(E_\textrm{p})=1$.
The distribution $n(E_\textrm{p})$ is shown in Fig.~\ref{Figure:Edist}(d) for $E=\Vr>0$.
The good agreement between the numerical data (dots) 
and the analytic prediction (Eq.~(\ref{nV2D}); solid line) legitimates the micro-canonical 
statistical model above, and, consequently, the phase-space ergodicity hypothesis.
Using Eq.~(\ref{nV2D}), we find that the average potential energy
of a particle of energy $E$ in the disordered potential is
\be \label{Ep1D_2D}
\langle E_\textrm{p} \rangle = \Vr \left( 1 -  \frac{ E/\Vr }{ \me^{E/\Vr} -1} \right),
~~\Vr>0.~~~~~
\ee
Then, using the relation imposed by energy conservation,
$E=\langle E_\textrm{k} \rangle + \langle E_\textrm{p} \rangle$,
we find that the average kinetic energy is given by
\be \label{Ek1D_2D}
\langle E_\textrm{k} \rangle = \Vr \left( \frac{ E/\Vr }{ 1 - \me^{-E/\Vr}} - 1 \right),
~~\Vr>0.~~
\ee
In particular, in the limit $E \ll \Vr$, we have
$\langle E_\textrm{p} \rangle \simeq E/2$ and $\langle E_\textrm{k} \rangle \simeq E/2$.
This is easily interpreted: The particles are trapped in minima of the disordered potential,
which can be assimilated to local (2D) harmonic oscillators,
and the total energy is equally partitioned over the four degrees of freedom.
In the opposite limit, $E \gg \Vr$,
the energy of the particles exceeds the disordered potential,
the velocity field is nearly uniform
and the average potential energy equals the spatial average of the disordered potential.
We have $\langle E_\textrm{p} \rangle \simeq \Vr$
and $\langle E_\textrm{k} \rangle \simeq E-\Vr$.
Finally, Eq.~(\ref{Ek1D_2D}) is in excellent agreement with the asymptotic value
of the kinetic energy obtained in numerical calculations [see Fig.~\ref{Figure:Edist}(c)].
For a red-detuned speckle potential, we find a similar behavior.
The counter-parts of Eqs.~(\ref{nV2D}), (\ref{Ep1D_2D}) and (\ref{Ek1D_2D})
have been calculated~\cite{note_redspeack} and checked numerically.

\subsection{Energy-dependent transport regimes}
\label{DiffusionRegimes}

In the following we study the mean-square displacement,
$\big\langle r(t)^2 \big\rangle$, along the $r \in\{x,y\}$ direction.
Note that, after disorder averaging, the two parity
symmetries $x \leftrightarrow -x$ and $y \leftrightarrow -y$
of the 2D (anisotropic) speckle potential impose
$\langle x(t) \rangle = \langle y(t) \rangle = \langle x(t)y(t) \rangle = 0$,
for $t \gg \tau^*(E)$ when fixing the initial momentum direction.
The above mean values hold at any time when averaging over an isotropic initial momentum distribution.

Figures~\ref{Figure:Results}(d)-(f) show the typical behavior of $\left\langle r(t)^2 \right\rangle$
as a function of time, for three different energies $E$ in a blue-detuned speckle potential ($\Vr>0$).
Quite generally, the plots in log-log scale show straight lines for long time periods.
In each of them, the dynamics is well described by the general formula
\be \label{rt}
\frac{\big\langle r(t)^2 \big\rangle}{\sigmar^2} = 2 \widetilde{D}_r \times \Big(\frac{t}{\taur}\Big)^{\alpha_r},
\ee
where the parameters $\widetilde{D}_r$ and $\alpha_r$ depend on the rescaled energy $E/\Vr$.
The different transport regimes are determined by the value of the exponent $\alpha_r$:
(i)~classical localization ($\alpha_r=0$), where $\big\langle r(t)^2 \big\rangle$ is independent of time;
(ii)~sub-diffusion ($0<\alpha_r<1$); 
(iii)~normal diffusion ($\alpha_r=1$);
(iv)~super-diffusion ($1<\alpha_r<2$); 
(v)~ballistic expansion ($\alpha_r=2$), which holds for free particles.

As discussed above, for times $t \lesssim \tau^*(E)$,
the dynamics is isotropic and ballistic.
We have
\be \label{ballistic}
\frac{\big\langle r(t)^2 \big\rangle}{\sigmar^2} = \overline{\mathrm{v}}_0^2 \times \Big(\frac{t}{\taur}\Big)^2,
\ee
where $\overline{\mathrm{v}}_0 = \sqrt{2 \langle E_\textrm{k} \rangle_0 /\vert\Vr\vert}$
and $\langle E_\textrm{k} \rangle_0$ is the kinetic energy
averaged over an isotropic momentum distribution.
For the 2D geometry considered here
we have $\langle E_\textrm{k} \rangle_0=\langle E_\textrm{k} \rangle$,
which for a blue-detuned speckle potential is
given by Eq.~(\ref{Ek1D_2D})~\cite{note_Ek0}.
Equation~(\ref{ballistic}) is the dotted black line in
Figs.~\ref{Figure:Results}(d)-(f)
and, as expected, it shows a very good agreement with the numerics \cite{note_Ek0} for short times.

For longer times ($t \gg \tau^*(E)$), the dynamics is strongly affected
by the disorder and we have $\alpha_r < 2$.
In the following, we discuss the {transport regimes} accessible to the particles,
depending on their energy.

\begin{figure}[!t]
\begin{center}
\includegraphics[scale=0.25]{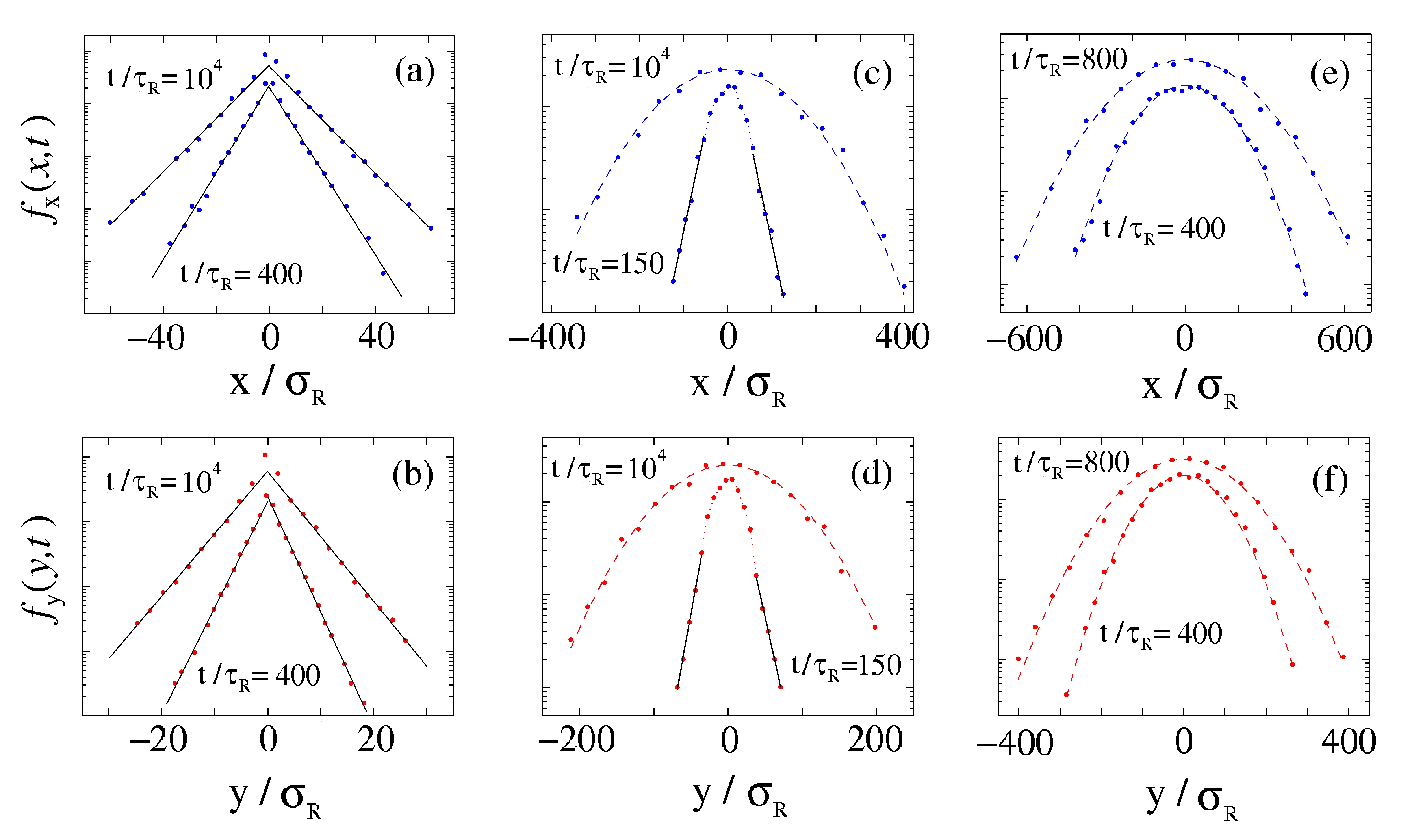}
\end{center}
\caption{\small{(color online) Integrated, average spatial distribution (dots)
along the $x$ (upper row) and $y$ (lower row) 
directions in semi-log scale.
Different sets of data in each figure refer to different evolution times,
vertically shifted for clarity and explicitly indicated.
Here we consider a blue-detuned speckle potential.
In panels (a) and (b) the energy is $E=0.3\Vr$.
The solid lines are fit to the function
$f_r(r,t) \approx e^{-b_r(t) |r|}$ which reproduces quite well the
numerical results in the tails of the distributions.
In particular, we have $b_x(t/\taur=10^4)=0.11/\sigma_R$ and
$b_y(t/\taur=10^4)=0.22/\sigma_R$, the ratio between the two
values being equal to the anisotropy factor $\lambda=2$.
In panels (c) and (d) the energy is $E=\Vr$.
The dotted lines are guides to the eye.
The black lines are fits with exponential functions, $f(r,t) \approx e^{-b_r(t) |r|}$,
for relatively short times ($t/\taur=150$), while the dashed lines
are $f(r,t) \approx e^{-r^2/4 D_r(E)t}$, reproducing the data for
long times.
In panels (e) and (f) the energy is $E=3\Vr$ and
the dashed lines are $f(r,t) \approx e^{-r^2/4 D_r(E)t}$.
The diffusion coefficients $D_r(E)$ are given 
by Eqs.~(\ref{DiffusionFit}) and (\ref{DiffusionFitBlue}).}}
\label{Figure:SpaceDist}
\end{figure}

\subparagraph{Classical localization regime.}
Let us {first} discuss the dynamical behavior of a particle of energy
below the percolation threshold ($E<E_\textrm{c}$).
{Figure~\ref{Figure:Results}(d) shows that, in average,
$\big\langle r(t)^2 \big\rangle$ is well described by Eq.~(\ref{rt}),
with different parameters in different time intervals.
At intermediate times,
the particle chaotically explores the finite-size allowed region it belongs to,
bouncing between the borders.
Its probability distribution progressively fills this region
and we then find a subdiffusive behavior ($\alpha_r<1$).}
At asymptotically large times, when the probability distribution has filled
all the allowed region, we find $\alpha_r=0$, corresponding 
to classical localization. 
In this case, the coefficients $\tilde{D}_r$ in Eq.~(\ref{rt}) are given
by the average mean-square size of the allowed regions
along the $x$ and $y$ directions [these are plotted as horizontal 
solid lines in Fig.~\ref{Figure:Results}(d)].
In an anisotropic disorder, we find that
$\big\langle x^2 \big\rangle / \big\langle y^2\big\rangle \simeq \lambda^2$,
which is consistent with the fact that the allowed regions are anisotropic,
with the same anisotropy factor as the potential.

In order to get further insight on the relation between
this behavior and the topography of the allowed regions,
we have studied the ``probability of diffusion'', $f(\vecr,t \vert \vecr_0, t_0)$,
which is the probability distribution to find a particle at position $\vecr$ at time $t$
assuming that it is at position $\vecr_0$ at time $t_0$.
Due to time translation invariance
and space transition invariance (after averaging over disorder configurations),
it is sufficient to consider the case $\vecr_0=0$ and $t_0=0$.
Numerical results for the {reduced probabilities
$f_x(x,t) = \int \ud y f(\vecr,t\vert\vecr_0, t_0)$
and $f_y(y,t) = \int \ud x f(\vecr,t\vert\vecr_0, t_0)$}
are shown in Fig.~\ref{Figure:SpaceDist}(a) and (b) 
for particles of energy $E=0.3\Vr$ in a blue-detuned speckle potential.
The distributions along $x$ and $y$ present tails that
can be well fitted by exponential functions.
This behavior can be quantitatively related to the probability distribution
of the sizes $\mathscr{L}$ of the allowed regions for a given energy $E$,
which is exponentially small,
$P_E(\mathscr{L}) \propto \exp(-a_E \mathscr{L}/\sigmar)$ (see Sec.~\ref{topology}).
Consider particles trapped in lakes (allowed region)
of linear length $\mathscr{L}$ along $x$.
They chaotically explore the lakes and the linear density along $x$
(averaged over the disorder) is almost uniform and scales as $1/\mathscr{L}$.
Hence, the ``probability of diffusion'' scales as
$f_x (x,t \rightarrow \infty) \sim \int_x^\infty d\mathscr{L}\ P_E(\mathscr{L})\frac{1}{\mathscr{L}}$.
We then find that, for $x \gg \sigmar / a_E$,
$f_x (x,t \rightarrow \infty)$ decays exponentially with the same characteristic
length of $P_E(\mathscr{L})$,
namely $\ln [f_x (x,t \rightarrow \infty)] \sim -a_E x / \sigmar$.
For $E=0.3\Vr$, we indeed find that, in the long time and long distance limits,
the decay rate of $f_x (x,t)$
[$b_x \simeq 0.11/\sigmar$; see Fig.~\ref{Figure:SpaceDist}(a)]
is approximately the same as that of $P_E$
[$a_E \simeq 0.12/\sigmar$; see the red triangles in Fig.~\ref{Figure:Loc}(a)].
The decay of $f_y(y,t)$ [see Fig.~\ref{Figure:SpaceDist}(b)]
is simply related to that of $f_x(x,t)$ by taking into account the 
anisotropy factor $\lambda=\sigma_x/\sigma_y$ of the disorder.
Therefore, the function $f_y(y,t)$ has exponential tails which decay $\lambda$
times faster than $f_x(x,t)$.

\begin{figure}[!t]
\begin{center}
\includegraphics[scale=0.34]{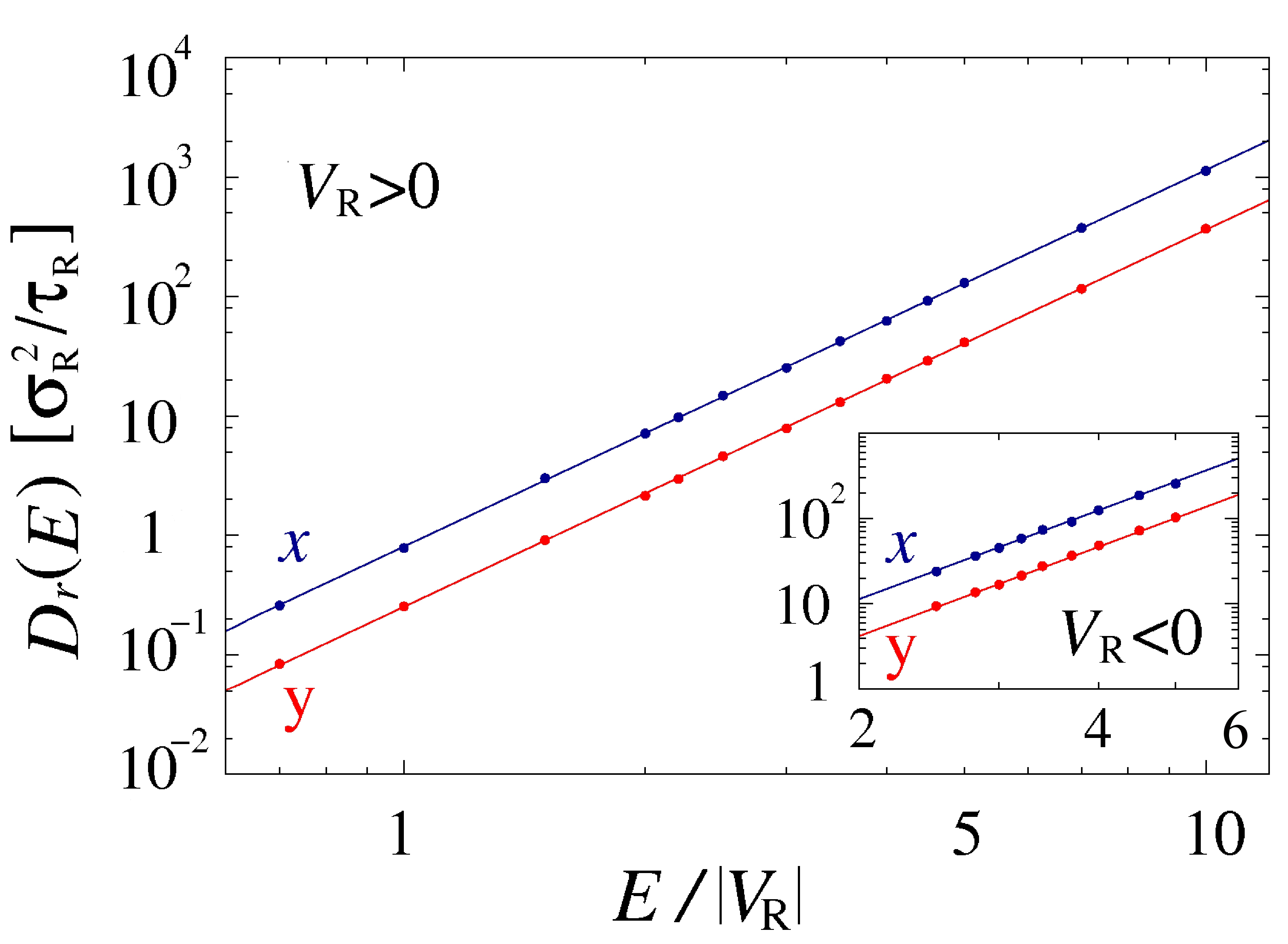}
\end{center}
\caption{\small{(color online) Diffusion coefficients $D_r(E)$ 
(dots; in units $\sigmar^2/\taur$) along the $r=x,y$ directions.
The results have been obtained numerically 
from fits of $\big\langle r(t)^2 \big\rangle$,
according to Eq.~(\ref{rt}), for large times.
The main figure shows the results for a blue detuned speckle.
Lines are the power-law fit with Eq.~(\ref{DiffusionFit})
and coefficients given by Eq.~(\ref{DiffusionFitBlue}).
The inset shows the diffusion coefficients $D_r(E)$ (dots)
for a red-detuned speckle disorder, as a function of 
$E/\vert V_R \vert +2$.
Lines are a power-law fit with Eq.~(\ref{DiffusionFit_red}) and
coefficients given by Eq.~(\ref{DiffusionFitRed}).
Numerical results have been obtained after averaging over 2000
realizations of the disordered potential.}}
\label{Figure:DiffCoeff}
\end{figure}

\subparagraph{Normal diffusion regime.}
For sufficiently long times and for energies above the
percolation threshold, the dynamics is
diffusive ($\alpha_r=1$).
This is shown in Fig.~\ref{Figure:Results}(e) and (f) where 
the solid lines are fits of
$\big\langle r(t)^2 \big\rangle = 2 D_r(E/\Vr) \times t$
to the numerical data for $t\gg\tau^*$, with $D_r$ as fitting parameter 
(along the $r=x,y$ directions).
We have studied the behavior of the diffusion coefficients $D_r(E/\Vr)$
as a function of the particle energy, for blue- and red-detuned speckle potentials.

\begin{figure}[!t]
\begin{center}
\includegraphics[scale=0.58]{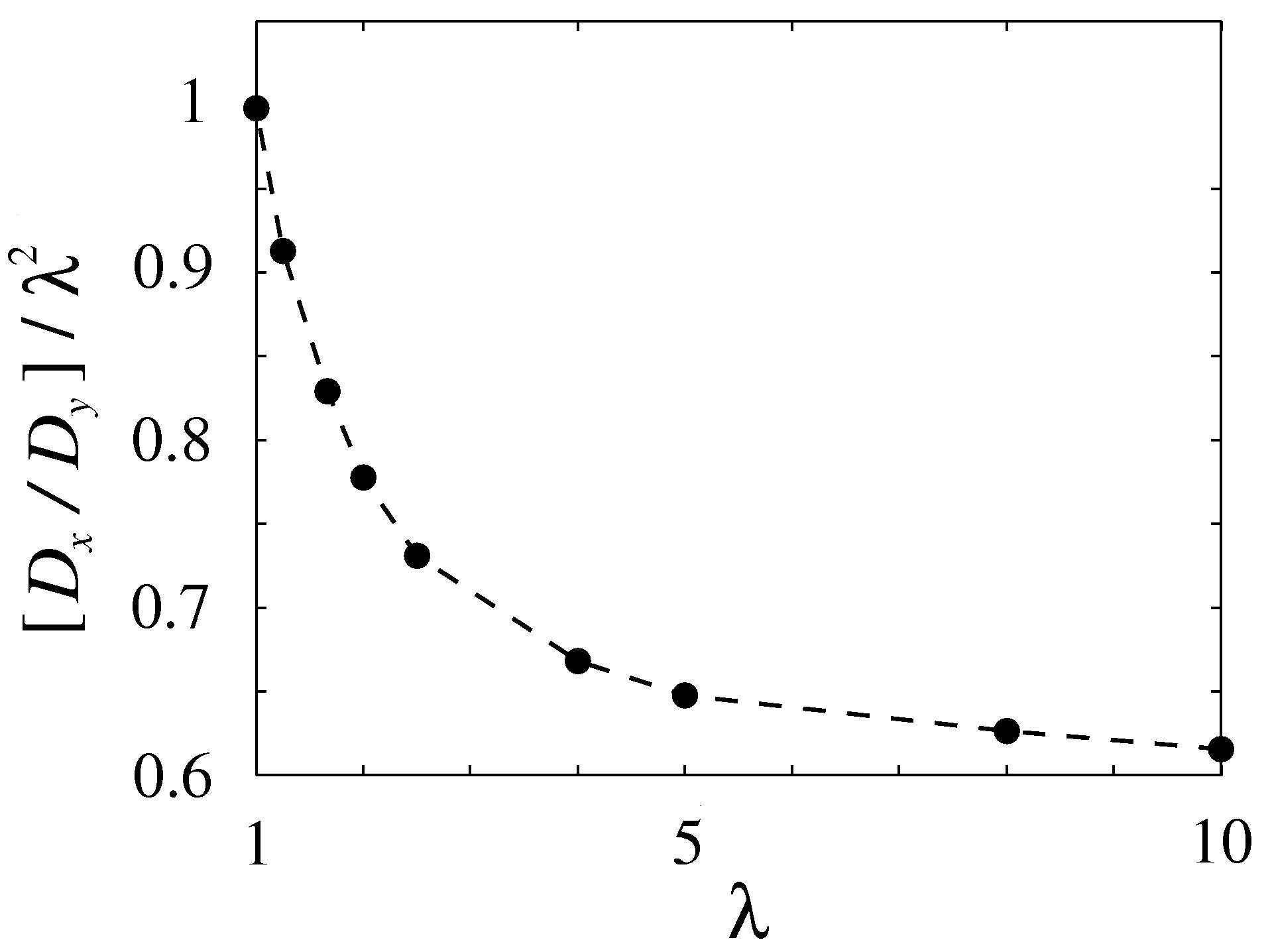}
\end{center}
\caption{\small{Anisotropy of the diffusion coefficient.  
Ratio $[D_x/D_y]/\lambda^2$ as a function of
$\lambda = \sigma_x / \sigma_y$.
Diffusion coefficients are obtained from the fit of 
$\big\langle r(t)^2 \big\rangle$, according to Eq.~(\ref{rt})
and for asymptotically large times.
Dots are obtained for $E=3\Vr$ and by averaging over 2000 realizations of 
a blue-detuned speckle potential.
The dashed lines is a guide to the eye.}}
\label{Figure:Anysotropy}
\end{figure}

For a blue-detuned speckle potential, we find a clear power-law scaling
{of the diffusion coefficients versus the energy,
\be \label{DiffusionFit}
D_r(E/\Vr) = D_r^0 \times \left(\frac{E}{\Vr}\right)^{\gamma_r},~~\Vr>0,
\ee
{as evidenced by Fig.~\ref{Figure:DiffCoeff},
which shows $D_r$ versus $E$ in log-log scale.}
Fits of Eq.~(\ref{DiffusionFit}) to the numerical data,
with $D_r^0$ and $\gamma_r$ as fitting parameters provide}
\begin{subequations} \label{DiffusionFitBlue}
  \begin{gather}
    D_x^0 \simeq (0.81\,\pm \,0.03) \,\, \sigmar^2/\taur, ~~~\Vr>0, \label{DiffusionFitBlue.1} \\
    D_y^0 \simeq (0.26\,\pm \,0.01) \,\, \sigmar^2/\taur, \phantom{~~~\Vr>0} \label{DiffusionFitBlue.2} \\
    \gamma_x \simeq \gamma_y \simeq 3.15\,\pm \,0.04. \phantom{~~~\Vr>0} \label{DiffusionFitBlue.3}
  \end{gather}
\end{subequations}
The diffusion coefficients $D_r$
are characterized, within numerical error,
by the same exponent $\gamma$ along the $x$ and $y$ directions.
In particular, the anisotropy of the diffusion
does not depend much on the energy \cite{note_AnisoFact}.
Interestingly, we find that, in the diffusion regime,
the anisotropy factor of the diffusion,
$D_x^0/D_y^0 \simeq 3.1$, significantly differs from the squared
anisotropy factor of the disordered potential ($\lambda^2 = 4$).
It shows that, differently from the classically localized regime,
the anisotropy of the diffusion differs from the 
anisotropy of the disorder correlation function.
This is further investigated in Fig.~\ref{Figure:Anysotropy},
which shows the ratio $(D_x/D_y)/\lambda^2$
as a function of $\lambda= \sigma_x / \sigma_y$
for $E=3\Vr$.
For $\lambda>1$, we find that $D_x(E)/D_y(E) < \lambda^2$.

For a red-detuned speckle potential, we have
\be \label{DiffusionFit_red}
D_r(E/\Vr) = D_r^0 \times \left(\frac{E-2\Vr}{\vert\Vr\vert}\right)^{\gamma_r}, ~~\Vr<0.
\ee
Fits of Eq.~(\ref{DiffusionFit_red}) to the numerics with $D_r^0$ and $\gamma_r$
as fitting parameters are shown in the inset of Fig.~\ref{Figure:DiffCoeff}.
We find
\begin{subequations} \label{DiffusionFitRed}
  \begin{gather}
    D_x^0 \simeq (1.19\,\pm \,0.07) \,\, \sigmar^2/\taur, ~~~\Vr<0, \label{DiffusionFitRed.1} \\
    D_y^0 \simeq (0.37\,\pm \,0.02) \,\, \sigmar^2/\taur, \phantom{~~~\Vr<0} \label{DiffusionFitRed.2} \\
    \gamma_x \simeq \gamma_y \simeq 3.45\,\pm \,0.09. \phantom{~~~\Vr<0} \label{DiffusionFitRed.3}
  \end{gather}
\end{subequations}
Compared to the blue-detuned case, we obtain
significantly larger
coefficients $D_{r}^0$ and larger exponents $\gamma_r$.
However, the anisotropy factor is approximately unchanged,
$D_x^0/D_y^0 \simeq 3.2$.

The power law scaling of the diffusion coefficients versus the 
particle energy in large energy windows is a rather general feature
of particle diffusion problems \cite{Kuhn_2005, Kuhn_2007, Miniatura_2009}.
Here, we find that, remarkably, the exponents $\gamma_{x,y}$ are about 
the same along the two directions.
For the energy window we are considering here, which is relevant to 
the experiment of Ref.~\cite{Robert_2010}, the disorder is strong.
This is suggested by the fact that the diffusion coefficients at a given energy are significantly different for blue- and red-detuned
speckle potentials.
For higher energies, the theory of quantum diffusion shows that the diffusion coefficients,
calculated in the Born approximation, only depend on the autocorrelation function of the disordered potential \cite{Rammer_book},
which is the same for the blue- and red-detuned speckle potentials used in the present study.
For \textit{isotropic} Gaussian correlation functions, one then finds $\gamma = 2.5$
\cite{Miniatura_2009, Beilin_2010}.
Note however that the latter result holds for $E \gg \Vr^2/\Er$, which can be equivalently written as $(E/\Vr)^2 \gg (k\sigma_{x,y})^2$.
The parameters used in the present work do not fulfill this condition since the 
validity of Newton's equation of motion requires $k\sigma_{x,y} \gg 1$
and our calculations are limited to $E \lesssim 10\vert\Vr\vert$.

In the normal diffusion regime, we expect that the
{probabilities of diffusion} are given by Gaussian functions \cite{Bouchaud_1990}.
In Fig.~\ref{Figure:SpaceDist}(e) and (f), we plot the (non-normalized) probability
distributions along the $x$ and $y$ directions, taken at different times
$t=400\taur$ and $t=800\taur$, for a blue-detuned speckle potential
{and for particles initially at the origin, $\vecr_0=0$}.
The numerical findings are in very good agreement with the 
functions $f_r(r,t) \propto e^{-r^2/4 D_r(E)t}$,
where the coefficients $D_r(E)$ are given by Eqs.~(\ref{DiffusionFit})
and (\ref{DiffusionFitBlue}).
We have checked that this holds for different energies in the normal diffusion regime.

{Finally, we have studied the correlation functions 
$c_k(E,t) = \langle x^k(t) y^k(t) \rangle/\langle x^k(t)\rangle \langle y^k(t) \rangle$
for various integer values of $k$.
For particles in the diffusion regime and for $t \gg \tau^*(E)$,
we find $c_k(E,t) \to 1$ for different values of $k$.
This shows that, in the normal diffusion regime,
the $x$ and $y$ variables are independent.}
Then, taking into account the results of Figs.~\ref{Figure:SpaceDist}(e) and (f),
the probability of diffusion, i.e.\ the probability density
to find a particle at coordinate $(x,y)$ which is in $\vecr_0=(x_0,y_0)$ at time $t_0$,
is given by
\beq \label{GaussApprox}
f_E(\vecr,t \vert \vecr_0,t_0) &=&
\frac{e^{-\frac{(x-x_0)^2}{4 D_x(E)(t-t_0)}}}{\sqrt{4 \pi D_x(E)}}
\frac{e^{-\frac{(y-y_0)^2}{4 D_y(E)(t-t_0)}}}{\sqrt{4 \pi D_y(E)}}
\frac{ \Theta(t-t_0) }{(t-t_0)}. 
\eeq

\begin{figure}[!t]
\begin{center}
\includegraphics[scale=0.35]{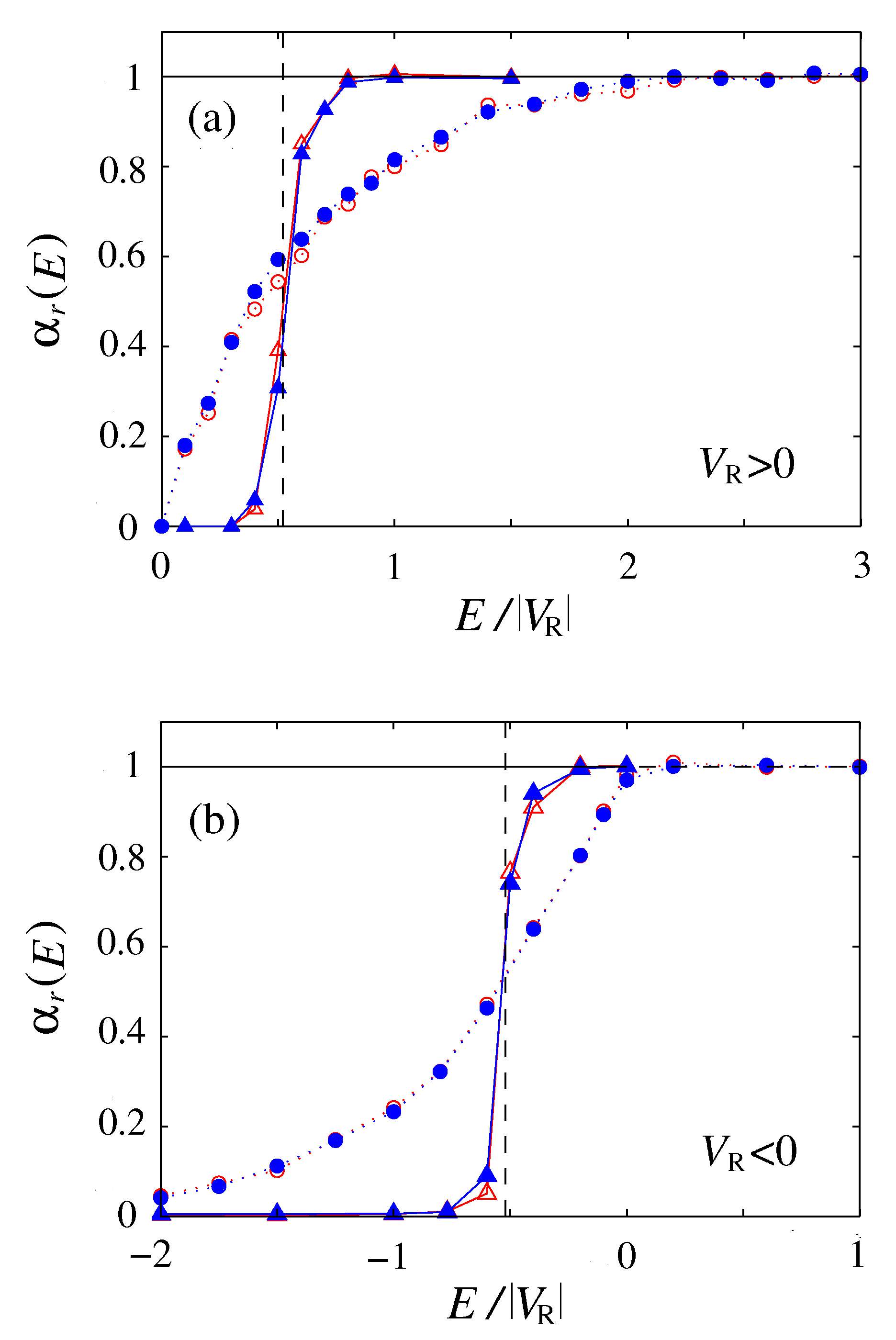}
\end{center}
\caption{\small{(color online) Scaling coefficient $\alpha_r(E)$ obtained from a fit of
$\big\langle r(t)^2 \big\rangle$ [according to Eq.~(\ref{rt})]
at intermediate times, $100 \lesssim t/\taur \lesssim 1000$,
(circles) and at larger times, $t/\taur \gtrsim 1000$, (triangles).
The fit are shown, for instance, in Fig.~\ref{Figure:Results}(d)-(f).
Solid and dotted lines are guides to the eye.
Different colors refer to $r=x$ (blue lines and filled symbols)
and $r=y$ (red lines and empty symbols) direction.
Panels (a) and (b) refer to blue- and red-detuned speckle potential, respectively.
In each panel, the vertical dashed black line is the percolation threshold,
$E_\textrm{c}=0.52 \Vr$.}}
\label{Figure:ScalingCoeff}
\end{figure}

\subparagraph{Sub-diffusion regime.}
Anomalous diffusion occurs in a variety of physical
systems \cite{Bouchaud_1990, Metzler_2000, Klages_book}
and has been experimentally studied
in Refs.~\cite{Bathelemy_2008, Mercadier_2009, Bardou_1994, Lucioni_2011} for instance.
It is defined by the condition $\alpha_r \neq 1$ {in Eq.~(\ref{rt})}
and two physically different regimes should be distinguished:
sub-diffusion when $0< \alpha_r < 1$
and super-diffusion when $1< \alpha_r < 2$.
In our case, super-diffusion only shows up for $t \sim \tau*$
in the crossover from ballistic expansion ($\alpha_r=2$ for $t \lesssim \tau^*$)
to normal or sub-diffusion (for $t \gtrsim \tau^*$).
Sub-diffusion is more relevant to our study.
For a blue-detuned speckle potential, a transient sub-diffusive
behavior is found for $E \lesssim 2\Vr$ and
long (although not asymptotic) time scales with $t \gg \tau^*$.
For instance, the fits of Eq.~(\ref{rt}) to $\big\langle r(t)^2 \big\rangle$
in Figs.~\ref{Figure:Results}(d) and (e),
at intermediate times and with $\widetilde{D}_r$ and $\alpha_r$ as fitting parameters,
(dashed lines) give $\alpha_r < 1$.
The scaling coefficients $\alpha_r$, as obtained from such fits
in two different time windows,
are plotted as a function of energy
in Fig. \ref{Figure:ScalingCoeff} for
blue-detuned (left panel) and red-detuned (right panel) speckle potentials.
The value of $\alpha_r$ obtained from the fits
in an intermediate time window ($10^2\taur<t<10^3\taur$)
smoothly crosses over from $\alpha_r=0$ to $\alpha_r=1$
when crossing the percolation threshold $E_\textrm{c}$.
For very long times ($t>10^3\taur$), the crossover is very sharp,
which is compatible with an abrupt transition
between two asymptotically-relevant dynamical regimes:
Below the percolation threshold ($E<E_\textrm{c}$),
the mean-square displacement $\big\langle r(t)^2 \big\rangle$ evolves
from subdiffusion to classical localization [see Fig.~\ref{Figure:Results}(e)].
Conversely, slightly above the percolation threshold ($E \gtrsim E_\textrm{c}$),
$\big\langle r(t)^2 \big\rangle$ evolves from subdiffusion to
normal diffusion [see Fig.~\ref{Figure:Results}(f)].
Notably, this case shows that the sub-diffusion regime
cannot always be considered a precursor of localization.

\begin{figure}[!t]
\begin{center}
\includegraphics[scale=0.25]{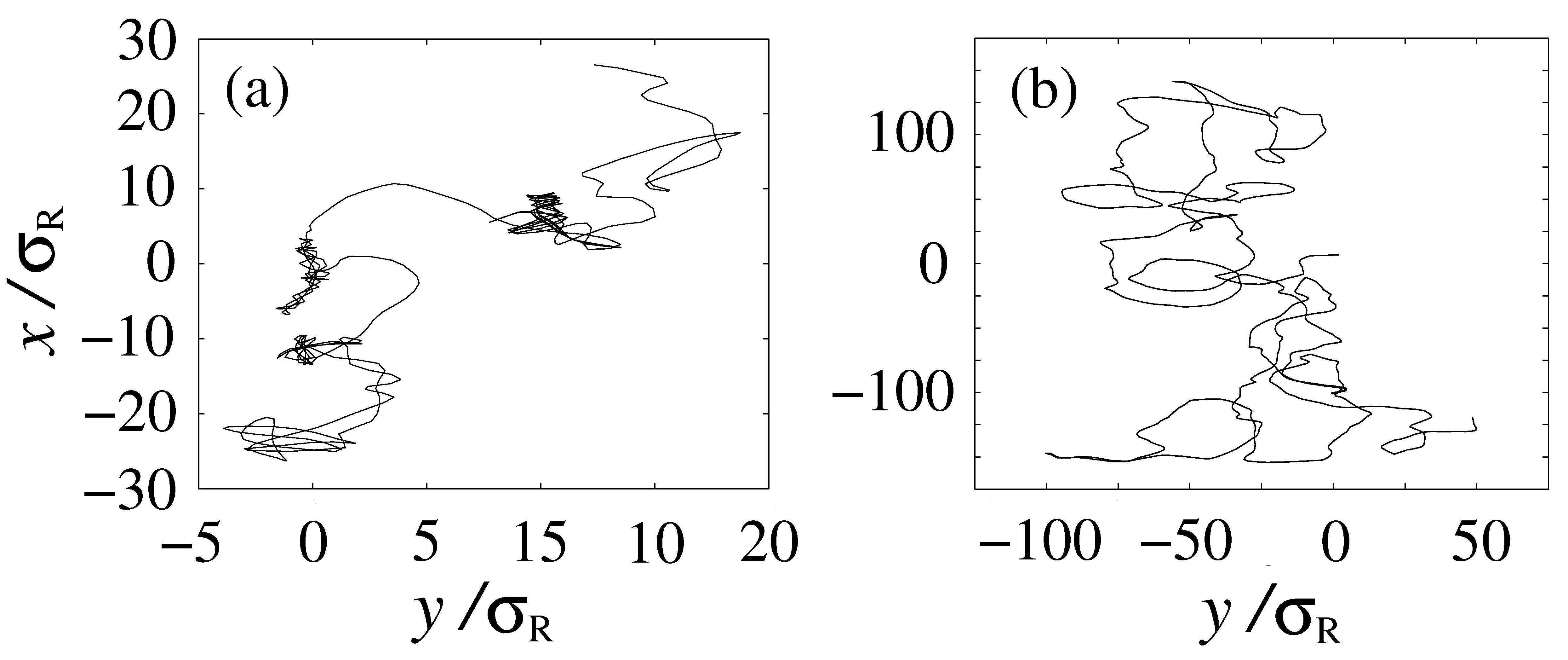}
\end{center}
\caption{\small{Trajectory of a particle of energy $E$
in an anisotropic, blue-detuned speckle potential, obtained by solving Eqs.~(\ref{ClEq}).
In the panel (a) the energy is $E=\Vr$ and the trajectory is
characterized by relatively small regions where the particle spends 
long times. This behavior give rise to transient sub-diffusion. 
The panel (b) shows the trajectory for a particle of energy $E=3\Vr$.
The dynamics is diffusive and qualitatively similar to a Brownian motion
\cite{Bouchaud_1990}.}}
\label{Figure:Traj}
\end{figure}

While the subdiffusive behavior is not asymptotic,
it is however experimentally relevant.
In fact, subdiffusion is observed for $E \sim \Vr$,
at times of the order of $\sim 200 \taur$,
which corresponds to $\sim 100$ms for the parameters of
Ref.~\cite{Robert_2010}, that is the typical time scale of the experiment.
In the sub-diffusion regime,
the probability density $f(\vecr,t \vert \vecr_0, t_0)$
is characterized by exponentially decaying tails
[see Figs.~\ref{Figure:SpaceDist}(a), (b), (c) and (d)].
We interpret this result by
analogy with the behavior described above in the classical localization regime.
Slightly above the percolation threshold, there exist allowed regions
of finite size that are disconnected from each other or connected by very thin bottlenecks
to a percolating cluster.
Then, as illustrated in Fig.~\ref{Figure:Traj}(a), a particle
stays long times in the same lake where it behaves similarly
as below the percolation threshold.
In particular, its dynamics is subdiffusive.
As discussed above, this persists for asymptotically large times 
when classical localization occurs
[for $E \leq E_\textrm{c}$; Figs.~\ref{Figure:SpaceDist}(a) and (b)].
Above the percolation threshold ($E>E_\textrm{c}$) however,
the particle can find a path connecting the lake to another one
after sufficiently long exploration of the lake.
It then spends a long time in the new lake before switching to another one,
and so on~\cite{noteDissip}.
For longer time scale and larger space scale,
the dynamics results from random jumps of the particle from a trapping region to another one,
hence developing normal diffusion, similarly as in usual Brownian motion.
In particular, the probability of diffusion becomes Gaussian
[see Figs.~\ref{Figure:SpaceDist}(c) and (d) for sufficient long time].
Here, the existence of long waiting times would explain
the transient sub-diffusive regime obtained at intermediate times.
Note that for higher energies, where no transient sub-diffusion is found,
the trajectory of a particle does not show long waiting times
and reminds that of Brownian motion even on a short times scale
[see Fig.~\ref{Figure:Traj}(b)].

\section{Expansion of an atomic cloud}
\label{AtomicCloud}

Section~\ref{FixedEnergy} presents a detailed analysis
of the transport regimes for particles with a fixed energy.
In typical experiments with cold atoms,
transport is rather investigated from the behavior of a cloud
of atoms with a broad energy distribution.
In this section, we study the expansion of an atomic gas by considering the atoms
as non-interacting classical particles moving in a 2D disordered potential.
We take into account the energy distribution of the gas and discuss our findings 
in the light of the results of Sec.~\ref{FixedEnergy}.

\subsection{Energy distribution of an initially trapped gas}

The aim of this section is to calculate the energy distribution of a 
thermal gas of non-interacting particles initially trapped. 
We will consider two experimentally relevant cases in which the 
disorder is initially combined with the trap or switched 
on after releasing the trap.
The energy distribution is highly sensitive to the way the disorder is turned on.

Quite generally, in transport experiments with cold atoms,  
the atomic cloud is initially in equilibrium in a
trap described by the potential $V_{0}(\vecr)$.
The initial single-particle classical Hamiltonian is
$H_{0}(\vecr,\vecp) = \vert \vecp \vert^2/2m + V_{0}(\vecr)$ and 
the phase-space density distribution for a thermal gas at temperature $T$ reads
\be
\rho_0(\vecr, \vecp) = \frac{e^{-\beta H_{0}(\vecr,\vecp)}}{Z_0},
\ee
where $Z_0 = \int \ud \vecr' \ud \vecp' \, e^{-\beta H_{0}(\vecr',\vecp')}$, 
$\beta = 1/k_\textrm{B} T$ and $k_\textrm{B}$ is the Boltzmann constant.
At time $t=0$, the confinement is switched off and the non-interacting
gas expands in the disordered potential $V(\vecr)$.
The dynamics is then governed by the single-particle Hamiltonian
$H(\vecr,\vecp) = \vert \vecp \vert^2/2m + V(\vecr)$
and the joint distribution of energy and position at time $t=0^+$ is
\beq 
n(\vecr, E) & = & \int \ud \vecp \,
\rho_0(\vecr, \vecp) \,
\delta \big( E -  H(\vecr,\vecp) \big) \\
& = & \frac{\beta \, e^{-\beta (E - V(\vecr) +V_0(\vecr) )}}{\int \ud \vecr' \, e^{-\beta V_0(\vecr')}} 
\Theta\big(E-V(\vecr)\big). \label{nrE0}
\eeq
In the following, we calculate the disorder-averaged 
energy distribution $\langle n(\vecr, E) \rangle$, 
distinguishing the two experimentally relevant situations:
{(i)~The gas is prepared in a bare trap,
so that $V_0(\vecr)=\Vtrap(\vecr)$, where $\Vtrap(\vecr)$ is the potential of the trap,
which is independent of the disordered potential. 
The disorder is suddenly switched on after the gas has thermalized in the trap. 
(ii)~The gas is prepared in the trap and in the presence of disorder,
so that $V_0(\vecr)=\Vtrap(\vecr)+V(\vecr)$,
which now depends on the disordered potential.}

\begin{figure}[!t]
\begin{center}
\includegraphics[scale=0.18]{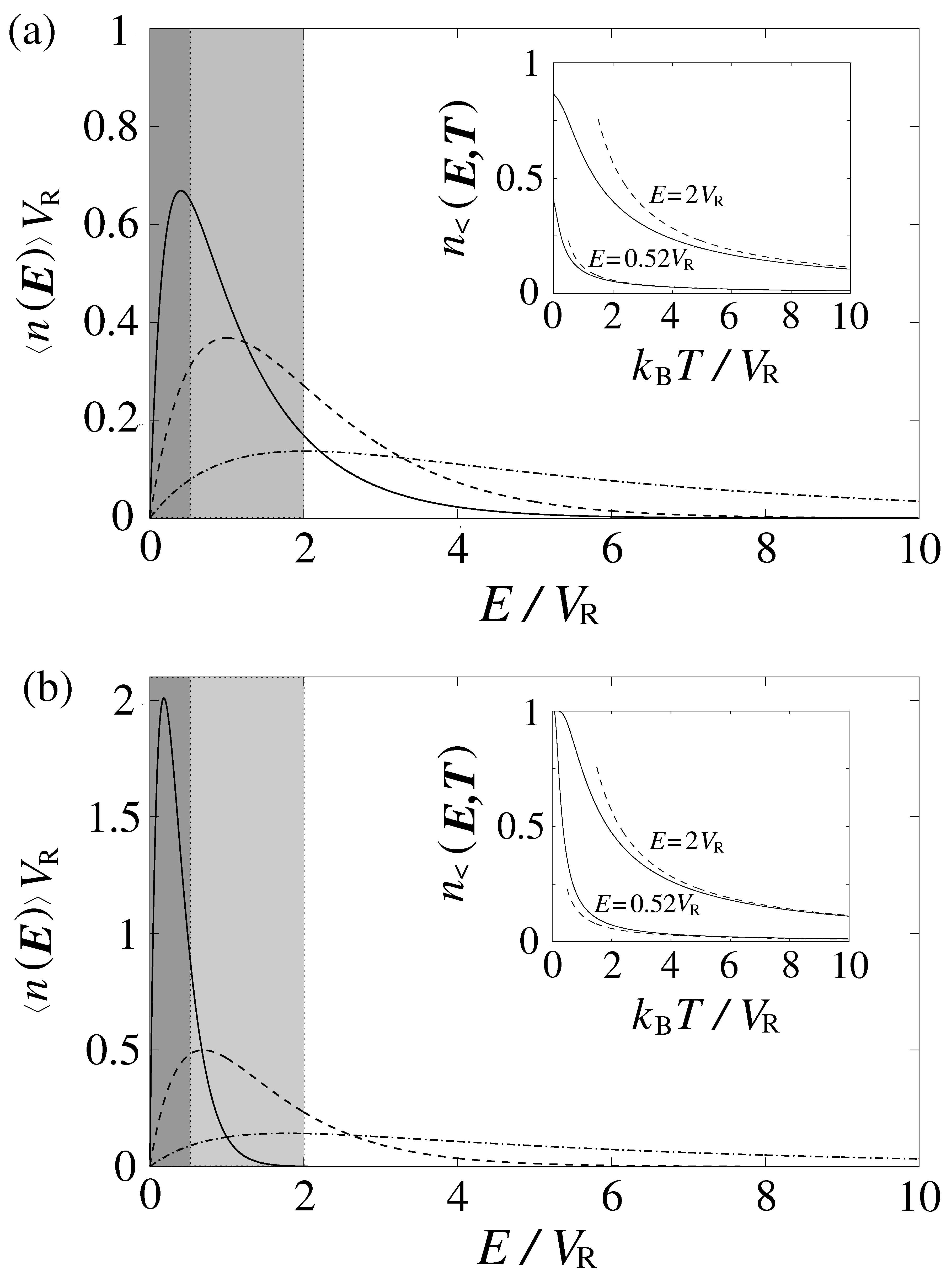}
\end{center}
\caption{\small{
Disorder-averaged energy distributions of the atomic cloud.
Panel (a) is Eq.~(\ref{NE1}), obtained when the speckle potential is
suddenly switch on after releasing the trap where the atomic cloud is initially in equilibrium [Case~(i)].   
Panel (b) is Eq.~(\ref{NE2}) and corresponds to a thermal cloud in equilibrium in a disordered trap [Case~(ii)].
The different curves refer to different values of $\Vr$:
$\Vr=5 \, k_\textrm{B} T$ (solid line), $\Vr= k_\textrm{B} T$ (dashed line) and $\Vr=0.2 \, k_\textrm{B} T$ (dot-dashed line).
The gray shaded region indicates
sub-diffusive particles ($E \leq 2 \, \Vr$), the darkest region 
classically localized particles ($E\leq 0.52 \, \Vr$) [see Sec.~\ref{FixedEnergy}].
Inset: density of particles with energy smaller than $E$ (solid lines).
The dashed lines are a prediction in the large temperature limit $k_\textrm{B} T \gg \Vr$, 
see footnote \cite{nota_N<E}.}}
\label{Figure:EDist}
\end{figure}

\subparagraph{Case~(i).}
If $V_0(\vecr)$ does not depend on the disordered potential,
averaging the distribution Eq.~(\ref{nrE0}) gives
\beq 
\langle n(\vecr, E) \rangle 
= \frac{\beta \, e^{-\beta V_0(\vecr)}}{\int \ud \vecr' \, e^{-\beta V_0(\vecr')}} 
\int_{-\infty}^{E} \ud V \, P(V) \, e^{-\beta ( E - V)},
\nonumber
\eeq
where $P(V)$ is the one-point probability distribution of the disordered potential.
It is interesting to notice that, in a homogeneous disorder and
independently of the specific form of the initial trapping potential, the energy and position 
variables decouple:
\be \label{decouple}
\langle n(\vecr, E) \rangle = \ntrap(\vecr) \langle n(E) \rangle,
\ee
where
\be \label{InitDensity}
\ntrap(\vecr) = \int \ud \vecp \rho_0(\vecr,\vecp)=
\frac{e^{-\beta V_0(\vecr)}}{\int \ud \vecr' e^{-\beta V_0(\vecr')}}
\ee
is the initial spatial distribution and 
\be \label{<nE1>}
\langle n(E) \rangle = \beta \, \int_{-\infty}^{E} \ud V \, P(V) \, e^{-\beta ( E - V)}
\ee
is the average energy distribution in the presence of disorder.
For a speckle potential, Eq.~(\ref{PV}) holds and the integral in Eq.~(\ref{<nE1>})
can be easily done.
In particular, for a blue-detuned speckle potential, we find \cite{note_NEred} 
\be \label{NE1}
\langle n(E) \rangle= \frac{\beta}{1-\beta \Vr} \Big[ e^{-\beta E} - e^{-E/\Vr}\Big] \, \Theta(E).
\ee
A plot of Eq.~(\ref{NE1}) as a function of the energy 
is shown in Fig.~\ref{Figure:EDist}(a) for different values of $\beta \Vr$.  

\subparagraph{Case~(ii).}
{If on the contrary, $V_0(\vecr)$ depends on the disordered potential,}
the calculation of Eq.~(\ref{nrE0}) is difficult
since both the numerator and the denominator 
depend on the specific disorder configuration.
A simplified expression is obtained if the thermal gas extends, 
in the initial trapping potential $V_0(\vecr)$, over a region 
much larger than the correlation length of the disorder.
This condition can be fulfilled for a sufficiently shallow confining trap.  
In this case, the self-averaging properties of the disordered potential 
allow us to substitute
the integral $\int \ud \vecr' \, e^{-\beta V_0(\vecr')}$
by the disorder-averaged value
$\langle \int \ud \vecr' \, e^{-\beta V_0(\vecr')} \rangle = 
\int \ud \vecr' e^{-\beta V_T(\vecr')} \int \ud V P(V) e^{-\beta V}$. 
We thus find that, also in this case, the energy and position
variables decouple.
Then, Eq.~(\ref{decouple}) still holds,
with the same density distribution, given by Eq.~(\ref{InitDensity}),
but a different energy distribution,
\be \label{<nE2>}
\langle n(E) \rangle = \frac{\beta e^{-\beta E} \int_{-\infty}^{E} \ud V P(V)}{\int_{-\infty}^{+\infty} \ud V P(V) e^{-\beta V}}.
\ee
In particular, for a blue-detuned speckle potential, we have \cite{note_NEred} 
\be \label{NE2}
\langle n(E) \rangle= \beta (1+\beta \Vr) e^{-\beta E} \Big[ 1- e^{-E/\Vr} \Big] \, \Theta(E).
\ee
In the case of an isotropic harmonic trap
$\Vtrap(\vecr)=m \omega^2 \vert \vecr \vert^2/2$, 
we find $\ntrap(\vecr) = e^{-\vert \vecr \vert^2/2 \sigma_T^2}/2 \pi \sigma_T^2$,
where $\sigma_T = \sqrt{k_\textrm{B} T / m \omega^2}$.
The self-averaging condition invoked above is thus $\sigma_T \gg \sigma_R$.
A plot of Eq.~(\ref{NE2}) as a function of energy 
is shown in Fig.~\ref{Figure:EDist}(b) for different values of $\beta \Vr$.
Equations~(\ref{NE1}) and (\ref{NE2}) tend, in the high-temperature limit,
$k_\textrm{B} T \gg E, \Vr$, to $\langle n(E) \rangle = \beta (1- e^{-E/\Vr})$.
At low temperature, $k_\textrm{B} T \ll \Vr$, the two distributions
differ substantially, Eq.~(\ref{NE2}) being more strongly peaked
at small energies.

\subparagraph{}
The insets of Fig.~\ref{Figure:EDist} show
the fraction of particles with energy lower than $E$,
$n_<(E,T) \equiv \int_0^E \ud \epsilon \, \langle n (\epsilon) \rangle$ \cite{nota_N<E},
as a function of temperature and for various energies.
In particular it shows the fraction of atoms
in the classically localized regime, $n_<(E=0.52 \Vr,T)$,
and below the transient sub-diffusive regime threshold, $n_<(E=2 \Vr,T)$,
identified in Sec.~\ref{FixedEnergy}.
The dotted line is the asymptotic behavior for large temperatures,
$n_<(E,T) = (e^{-E/\Vr}+E/\Vr-1)\beta \Vr$,
which is {the same for both cases~(i) and (ii)}.
When the gas thermalizes in the trap
and in the presence of disorder [case~(ii) described by Eq.~(\ref{NE2})],
the fraction of particles at low energy is larger than in the case 
when the disordered is suddenly ramped after releasing the confining trap 
[case~(i) described by Eq.~(\ref{NE1})].
In case~(ii), all particles can be 
classically localized at low temperature [i.e. $n_<(E=0.52 \Vr, T) \to 1$ when $T \to 0$].
Conversely, in case~(i),
the speckle potential is switched on after releasing the initial confining trap, 
which transfers an average energy $\sim \Vr$ to all particles
and the fraction of particles at low energy is depleted.

\subsection{Expansion of the atomic gas}

We now study the dynamics of an atomic cloud
initially trapped in a pure harmonic potential
[i.e.\ without disorder, case~(i)].
At $t=0$ the trap is released and the atoms expand in 
a blue-detuned speckle potential.
As shown above, the energy and the initial position of 
the atoms are uncorrelated variables, when averaging over
disorder configurations.
In our simulations, the energy is randomly chosen according to Eq.~(\ref{NE1}), while
the initial position is randomly chosen according to a Gaussian distribution 
of width $\sigma_T=10\sigmar$, which is relevant to typical experiments \cite{Robert_2010}.
We expect this approach to be reproducible {\it in a single run of the experiment} since, 
for $\sigma_T \gg \sigmar$, the speckle disorder ``self-averages'' over the atomic cloud.

\subparagraph{Density distribution.}
An important feature of
the experiments with cold atoms is the possibility to obtain
an image of the density cloud at different expansion times.
For instance, it has been crucial in the recent observation
of Anderson localization with Bose-Einstein condensates 
in the quasi-1D geometry \cite{Billy_2008, Roati_2008}.

\begin{figure}[!t]
\begin{center}
\includegraphics[scale=0.37]{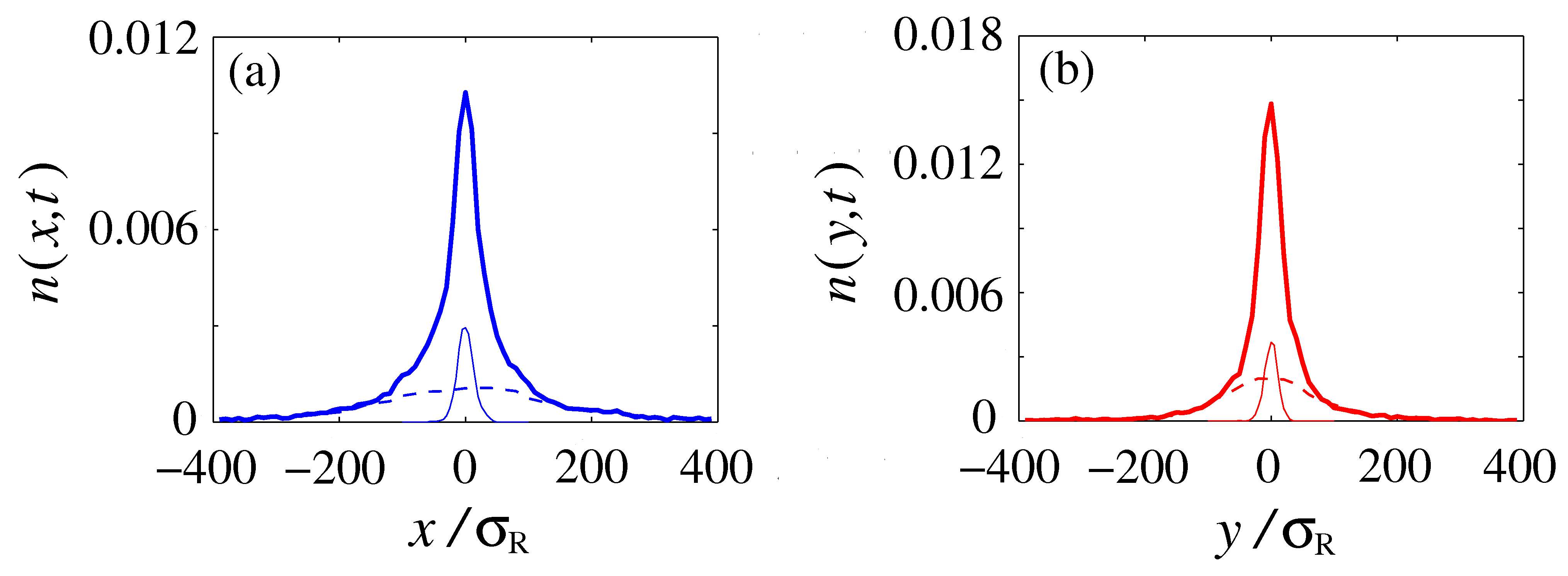}
\end{center}
\caption{\small{(color online) 
Integrated spatial density distribution of an expanding thermal gas 
in a 2D disordered potential.
The thick solid line is the integrated spatial density 
distribution
{(a)} $n_x(x,t) \equiv \int \ud y \, n(\vecr,t)$
and
{(b)} $n_y(y,t) \equiv \int \ud x \, n(\vecr,t)$,
where 
$n(\vecr,t)$ is given by Eq.~(\ref{fBEC}).
Here, $\Vr=k_\textrm{B} T$, $t=400\taur$, the initial distribution 
is a Gaussian of width $\sigma_T=10\sigma_R$ and the energy distribution is 
given by Eq.~(\ref{NE1}). In the numerical simulation, we used 10000 particles.
We also plot the contributions of particles with energy
$E>2 \Vr$ (dashed line) and with energy $E \leq 0.52 \Vr$ (thin solid line).
Note that for the time considered here, the transient sub-diffusive regime
is relevant and the fraction of atoms in this regime cannot be neglected.}}
\label{Figure:BECxyDist}
\end{figure}

The time-dependent density profile 
(averaged over different realizations of
the disordered potential) is calculated by integrating 
the probability of diffusion over all energies and all initial positions
\be \label{fBEC}
n(\vecr,t) = \int \ud \vecr_0 \int
\ud E \, \ntrap(\vecr_0) \, \langle n(E) \rangle \, f_E(\vecr, t \vert \vecr_0,0),
\ee
where $\ntrap(\vecr_0)$ is the spatial density distribution in the initial trap.
As noticed in the previous paragraph, for a thermal gas, 
the initial joint position-energy distribution decouples into the product of separate distributions.
This feature is particular to the case of an initial thermal distribution.
For instance, it does not hold for initially expanding Bose-Einstein condensates as considered in Refs.~\cite{Miniatura_2009,Piraud_2011,Shapiro_2007}.
The quantity $f_E(\vecr, t \vert \vecr_0,0)$ gives the probability distribution to
find a particle of energy $E$ at position $\vecr$ and at time $t \geq 0$,
assuming it was in $\vecr_0$ at time $t=0$.
Its behavior has been discussed in Sec.~\ref{FixedEnergy}
for particles of fixed energy in various transport regimes.
The integrated
{density profiles} at time
$t=400 \taur$ are shown as thick solid lines in Fig.~\ref{Figure:BECxyDist}.
{It is striking that these profiles
are in general, not Gaussian.}
They present a pronounced central peak,
which is partially due to the contributions of
the atoms in the classical localization and sub-diffusive regimes.
It is worth stressing that, even taking into account only
the fraction of atoms in the diffusive regime,
for which the probability of diffusion, $f_E(\vecr, t \vert \vecr_0,0)$,
is a Gaussian function [see Eq.~(\ref{GaussApprox})],
the overall distribution is, in general, not Gaussian.
This is due to the integration over the energy variable in Eq.~(\ref{fBEC}) 
and to the strong energy dependence of the diffusion coefficients \cite{Shapiro_2007,Beilin_2010}.
In fact, since the particles in the diffusive regime determine the long tails of $n(\vecr,t)$, 
the spatial distribution can be used to extract the diffusion coefficients
from experimental data (see Ref.~\cite{Robert_2010}
and Sec.~\ref{Exp}).

\begin{figure}[!t]
\begin{center}
\includegraphics[scale=0.55]{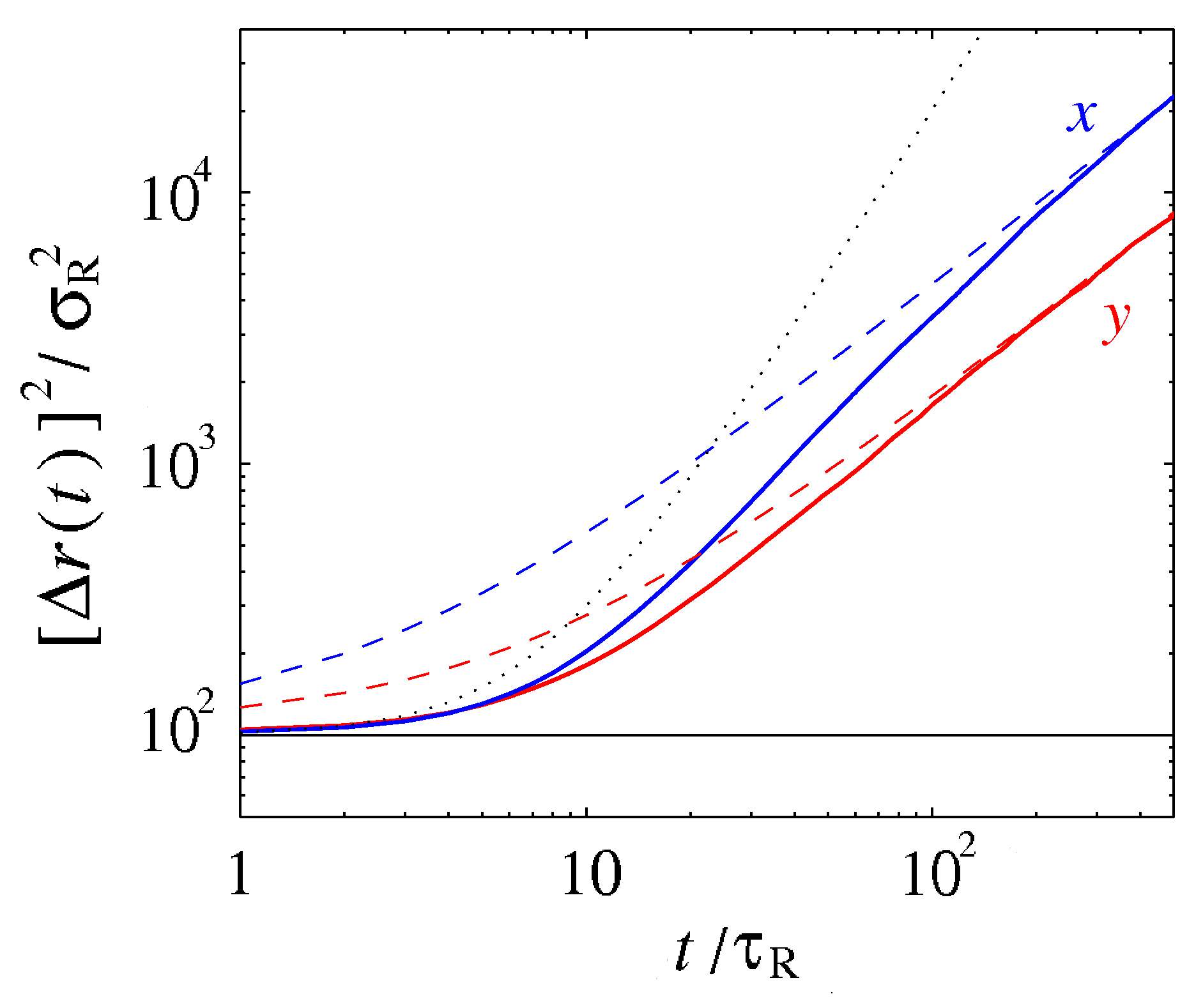}
\end{center}
\caption{\small{(color online) Mean square size, $[\Delta r(t)]^2$,  as a function of time,
in the $r=x$ (solid blue line) and $r=y$ (solid red line) directions.
The dotted line is Eq.~(\ref{ballisticMSF}) and has been obtained by assuming 
a ballistic expansion for all atoms.
The dashed lines are Eq.~(\ref{MSF}), corresponding to all particles behaving diffusively. 
The horizontal line is the mean square size of the initial cloud, given by $\sigma_T^2$.
Here $\Vr=k_\textrm{B} T$ and $\sigma_T =10 \sigma_R$.}}
\label{Figure:BECMSF}
\end{figure}

\subparagraph{Mean square size.}
A simple quantity which can be extracted from the density profiles
is the rms size of the atomic cloud,
$\Delta r(t) \equiv \sqrt{\int \ud \vecr \, r^2 \, n(\vecr,t)}$,
in the $r=x,y$ direction and where $n(\vecr,t)$ is given by Eq.~(\ref{NE1}).
{The evolution of the rms sizes in the two directions are shown
in Fig.~\ref{Figure:BECMSF}, for $\Vr=k_\textrm{B} T$.}
At sufficiently short times, $t \ll \tau^*(\bar E)$,
where $\bar E$ is the
{typical energy of the atomic cloud} \cite{note_barE},
the {rms size}
can be obtained by assuming that all the atoms behave ballistically \cite{note_ballistic}.
In this case, we find 
\be \label{ballisticMSF}
[\Delta r(t)]^2 = \sigma_T^2 + \frac{1+\beta\Vr}{\beta\Vr} \sigmar^2
\left( \frac{t}{\taur} \right)^2,
\ee
where
{$\sigma_T = \sqrt{k_\textrm{B}T/m\omega^2}$ is the rms size}
of the initial atomic cloud.
Equation (\ref{ballisticMSF}) is shown by the dotted black line in Fig.~\ref{Figure:BECMSF}.
For long times, $t \gg \tau^*(\bar E)$,
the rms size of the atomic cloud
can be calculated by assuming that all the particles behave diffusively.
This approximation is legitimate in the long-time limit
if the fraction of classically localized particles is negligible,
which strongly depends on the ratio $k_\textrm{B} T/\Vr$.
In this case, the evaluation of Eq.~(\ref{fBEC})
using Eqs.~(\ref{NE1}), (\ref{GaussApprox}) and (\ref{DiffusionFit}) gives
\be \label{MSF}
[\Delta r(t)]^2 \simeq
2 D_r^0 \sigmar^2
\frac{\big[1-(\beta \Vr)^{\gamma_r+1}\big]\Gamma(\gamma_r+1)}{(1-\beta\Vr)(\beta\Vr)^{\gamma_r}}
\frac{t}{\taur},
\ee
where $\Gamma(x)$ is the Gamma function.
In Fig.~\ref{Figure:BECMSF} we show Eq.~(\ref{MSF}) calculated for the parameters
$D_r^0$ and $\gamma_r$ given by Eqs.~(\ref{DiffusionFitBlue}).
We find a good agreement with the numerical findings for long times,
despite the fact that the fraction of localized or sub-diffusive
atoms is $n_<(E=2\Vr,T) \simeq 70\%$.
Hence, Eq.~(\ref{MSF}) could be used experimentally to extract the values of 
$\gamma_r$ and $D_r^0$ characterizing the diffusive regime,
and thus verify the results of Sec.~\ref{DiffusionRegimes}.

\section{Experimental results}
\label{Exp}

The study of classical transport of cold gases 
in a 2D disordered potential is experimentally 
feasible.
A 2D disordered potential can be created experimentally 
by a speckle field, as discussed in Sec.~\ref{SpeckleStat}.
The atoms can be confined to a plane
by a tight transversal harmonic trap of frequency $\omega_z$.
The dynamics in the transverse direction 
can be neglected if $k_\textrm{B}T$ is much smaller than the
vertical confinement energy $\hbar \omega_z$.

In recent experiments \cite{Robert_2010}, we have studied the 
dynamics of a cloud of $^{87}$Rb atoms in a speckle disorder. 
The geometry is not strictly 2D 
since $k_\textrm{B}T\approx k_\textrm{B} \times
200(20)\,$nK$\approx 6 \hbar \omega_z$, where 
$\omega_z/2\pi=680\,$Hz. 
Therefore, taking into account the harmonic confinement, 
the external potential is given by
$V(x,y,z)+m \omega^2 z^2/2$,
with $V(x,y,z) = V_{\mathrm{iso}}(x\sin \theta - z \cos \theta,y)$
where $V_{\mathrm{iso}}(u,v)$ is a 2D isotropic blue-detuned 
speckle potential with correlation length $\sigma=0.8$ $\mu$m
and Gaussian autocorrelation function.
In the experiment, the speckle disorder is tilted by
$\theta=30^{\mathrm{o}}$ with respect to the horizontal plane.
This makes the disorder anisotropic in the expansion plane, with correlation lengths
$\sigma_x=1.6$ $\mu$m and $\sigma_y=0.8$ $\mu$m, and anisotropy factor 
$\lambda=\sigma_x/\sigma_y=2$.
The speckle is assumed to be invariant along its propagation axis.
This is a reasonable approximation since the longitudinal correlation length,
$\sigma_{long}=9$ $\mu$m is large compared to the vertical size of the cloud,
$(k_\textrm{B}T/m \omega_z^2)^{1/2}\approx 1\,\mu$m.  

Using a classical particle model is a reasonable approximation
since $k\sigma_x  \approx 10$ and $k\sigma_y  \approx 5$, where $k=\sqrt{m k_B T}/\hbar$.
In addition, the ``correlation energy'' 
$E_\textrm{\tiny{R}}=\hbar^2/m\sigma_\textrm{\tiny{R}} 
\approx k_\textrm{\tiny{B}} \times 2$ nK
is typically more than one order of
magnitude smaller than the amplitude of the disorder 
$V_\textrm{\tiny{R}} = k_\textrm{\tiny{B}} \times 53$ nK.
Within these conditions, the localization length should be 
much larger than the system size ($\sim 1$ mm in each direction)
and one can expect that quantum interference effects 
can be neglected.
Due to the tilting angle of the speckle pattern, which
couples the in-plane dynamics to the dynamics in the confined
direction, the system is 3D rather than purely 2D. 
Therefore, the classical equations of
motions (\ref{ClEq}) must be adapted to take into account the
vertical confinement \cite{note3Deqs}. 

Remarkably, the results of this 3D model differ only slightly from the results
presented above for the purely 2D situation. 
The two asymptotic regimes
(classical localization and normal diffusion)
are unchanged since the 3D potential results from a 
cylindrical stretching of a 2D potential.
We also find a transient sub-diffusion regime in approximately the
same energy window and time scale as for the pure 2D case, $0.52\Vr \lesssim E \lesssim 2\Vr$.
For each regime, the dynamics is qualitatively similar
for both the pure 2D and the trapped 3D cases.
The diffusion coefficients have a power law scaling analogous to  
Eq.~(\ref{DiffusionFit}) \cite{Robert_2010}.
The numerical simulations give diffusion parameters slightly 
different to the ones of Eq.~(\ref{DiffusionFitBlue}):
\begin{subequations} \label{DiffCoeff3D}
  \begin{gather}
    D_x^0 = (1.20 \, \pm \, 0.05) \,\, \sigmar^2/\taur, \\
    D_y^0 = (0.33 \, \pm \, 0.02) \,\, \sigmar^2/\taur, \\
    \gamma_x \simeq \gamma_y = 2.8 \, \pm \, 0.06.
  \end{gather}
\end{subequations}
For the parameters of the experiment, $\Vr=k_\textrm{B} \times 53$ $n$K,
$\sigmar=\sigma_x=1.6$ $\mu$m and $\taur=0.71$ ms,
we obtain $D_x^0 = 4.3$ $\mu$m$^2$/ms and $D_y^0 = 1.2$ $\mu$m$^2$/ms.

\begin{figure}[t!]
\centering
\includegraphics[width=0.45\textwidth]{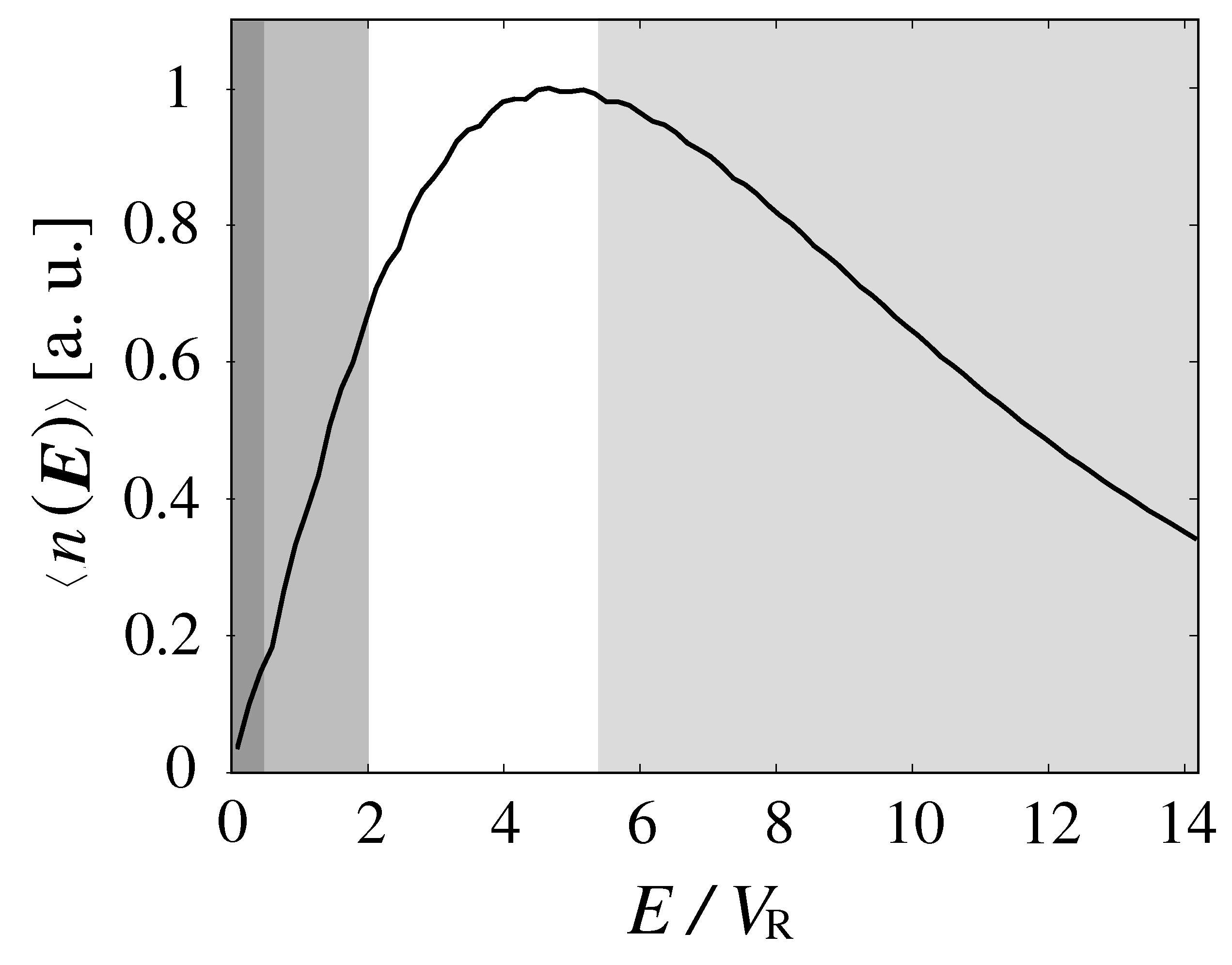}
\caption{Disorder-averaged energy distribution  
in the experiments of Ref.~\cite{Robert_2010}. 
The solid line is Eq.~(\ref{nEexp}).
Gray regions indicate different dynamical regimes: 
classical localization for $E<0.52\Vr$ (darkest region), 
transient subdiffusion for $E<2\Vr$ (gray region),
diffusion for $2\Vr<E<5.5\Vr$ (white region) 
and transient ballistic expansion (see text for details) for $E>5.5\Vr$ (light gray region). 
Here the amplitude of the disordered potential is 
$V_\textrm{R}=k_\textrm{B} \times 53\,$nK and 
the temperature $T = 220\,$nK.} \label{Figure:NEexp}
\end{figure}

In the experiment, the speckle disorder is switched on after releasing the 
initial confining trap in the ($x$,$y$) direction, while the transverse 
confinement is kept during the whole expansion.
We expect that a few collisions redistribute the energy between 
the atoms during the first 10 ms of the expansion.
The calculation of $\langle n(E) \rangle$ is
modified compared to Eq.~(\ref{NE2}) in order to include the
contribution of the different vertical energy levels. 
Taking into account the quantization of the transverse harmonic oscillator, we obtain
\begin{equation} \label{nEexp}
\langle n(E) \rangle \propto
\textrm{e}^{-E/k_\textrm{B} T} \sum_{n=0}^{E/\hbar \omega_z} \left(
1 - \textrm{e}^{(n \hbar \omega_z -E)/ \Vr} \right)
\end{equation}
which is plotted in Fig.~\ref{Figure:NEexp}. 
For the considered experimental parameters, the fraction of classically 
localized atoms is about $0.4\%$ of the total number of atoms
while the fraction of subdiffusive atoms for the time scale of the experiment
is about $6\%$.


\begin{figure}[t!]
\centering
\includegraphics[width=0.45\textwidth]{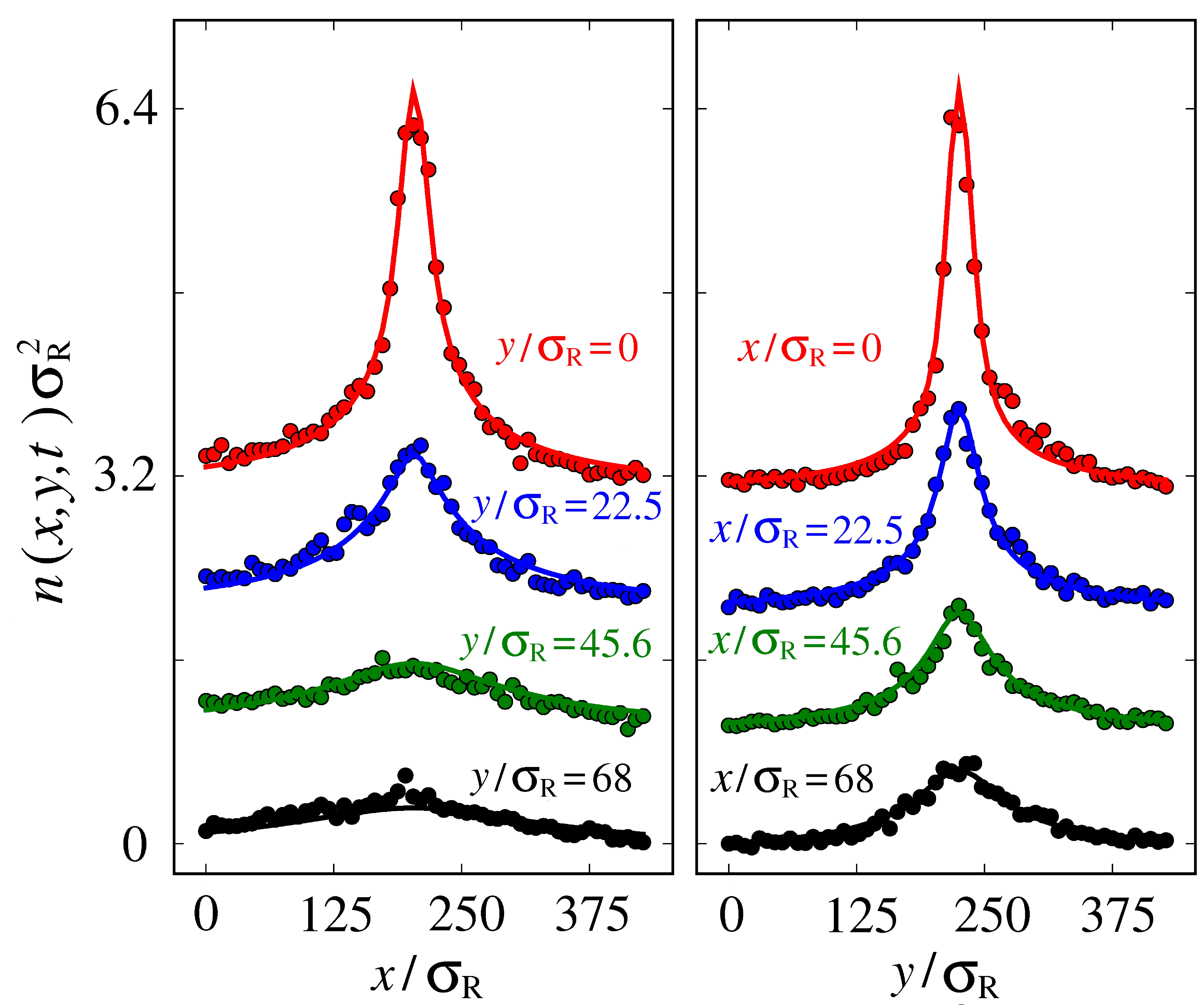}
\caption{(color online) Experimental two-dimensional 
column density distribution after 200\,ms ($=281 \taur$) expansion 
of the atomic cloud.
Dots are cuts along $x$
(left panel) and $y$ (right panel) directions at   
cut positions explicitly indicated, artificially offset for clarity.
Lines are the fitting function Eq.~(\ref{ExpFit}).
Note that the spatial distribution is not Gaussian.} \label{Figure:2Dcuts}
\end{figure}

The goal of the experiment is to extract the energy-dependent diffusion coefficients
in the 2D anisotropic speckle potential.
They are experimentally measured by analyzing the complete
2D {column} density profile, obtained experimentally by 
fluorescence imaging along the vertical axis (see Fig.~\ref{Figure:2Dcuts}). 
In order to fit the experimental density, we use Eqs.~(\ref{GaussApprox})
(with $t_0=0$) and (\ref{fBEC}) convolved by a Gaussian function which 
takes into account the 
resolution of the imaging system.
Using the fact that, for an initial harmonic trap, $\ntrap(\vecr_0)$ is a Gaussian 
function, we find 
\be \label{ExpFit}
n(x,y,t) = \int
\ud E \, \langle n(E) \rangle \, \frac{
e^{-\frac{x^2}{2 (2 D_x(E)t + \sigma_0^2)}} 
e^{-\frac{y^2}{2 (2 D_y(E)t + \sigma_0^2)}}}
{2 \pi \, t\, \sqrt{(2 D_x(E)t + \sigma_0^2)(2 D_y(E)t + \sigma_0^2)}},
\ee
where $\langle n(E) \rangle$ is given by Eq.~(\ref{nEexp})
and $\sigma_0=15$ $\mu$m as determined experimentally.
Equation (\ref{ExpFit}) assumes that all the particles behave 
diffusively. This is a good approximation since, for the experimental parameters, 
the fraction of classically localized and subdiffusive particles is small.
Diffusion coefficients are extracted by using the power law scaling 
predicted numerically, $D_r(E)=D_r^0 \times (E/V_R)^{\gamma_r}$ ($r=x,y$). 
Figure \ref{Figure:2Dcuts} shows that the 2D density distribution
is well reproduced by the fitting function, Eq.~(\ref{ExpFit}), 
with $D_r^0$ and $\gamma_r$ as
fitting parameters. Experimentally, we
find:
\begin{subequations} 
  \begin{gather}
    D_x^0 = 5.5 \pm 2.7 \,\, \mu \mathrm{m}^2/\mathrm{ms}, \\
    D_y^0 = 1.6 \pm 0.8 \,\, \mu \mathrm{m}^2/\mathrm{ms}, \\
    \gamma_x \simeq \gamma_y = 3.3.
  \end{gather}
\end{subequations}
These values are, within experimental errors,
in fair agreement with the classical theory [see Eqs.~(\ref{DiffCoeff3D})].
In particular, the algebraic scaling inferred from
the numerical simulations is consistent with the experimental
findings.
In addition, these results show the possibility
to extract the single-particle diffusion coefficients
from the dynamics of an atomic cloud.


We finally notice that the high energy atoms are
ballistic on the time scale of the experiment (see Fig.
\ref{Figure:NEexp}). The crossover between the ballistic and the diffusive behavior
takes place when the Boltzmann time $\tau_\textrm{B}$
($=4D(E)/v^2$, where $v=\sqrt{E/m}$ is the rms in-plane velocity)
is of the order of the expansion time $t=200$ ms. The same
crossover between ballistic and diffusive behaviors is seen as a
function of time in Fig. \ref{Figure:Results}(c) for a given
energy. For the experimental parameters, we estimate that below
$E_\textrm{bal-dif}\approx 5.5 V_{\textrm{R}}$, the atoms are
diffusive (calculated in the $x$ direction which gives the lower
estimate of $E_\textrm{bal-dif}$). Above this value, the atoms may
be ballistic {on the time scale of the experiment}.
In this case they however extend over a distance $vt
\gtrsim 1\,$mm, larger than the spatial size used in Fig.~\ref{Figure:2Dcuts}
where we fit our diffusive distribution. Ballistic atoms thus do
not affect the result of the fit for the diffusion coefficients.
Note that this is not the case after 50\,ms as ballistic atoms are
visible (see Fig.~2 in Ref. \cite{Robert_2010}).

In addition to the investigation of the normal diffusion
regime discussed above \cite{Robert_2010}, we have recently performed 
experiments with a lower temperature of 30\,nK.
In this case, Eq.~(\ref{nEexp})
leads to $8\%$ of classically trapped atoms and $55\%$ of atoms in
the subdiffusive regime. 
We observe little expansion of the cloud as a function
of time. 
For times of several seconds we observe losses due
to residual heating. 
We are unable to quantitatively analyze
the shape of the cloud because of the limited resolution of the
imaging system and of the initial size of the cloud. 
The subdiffusive regime and the percolation transition presented in
this manuscript thus remain to be experimentally studied in 2D
cold atomic systems.

\section{Conclusions}
\label{Conclusions}

In this work, we have studied the transport properties
of a cold atomic gas in a 2D anisotropic speckle potential,
in the classical regime.
We have shown that, upon proper rescaling, the atomic dynamics
can be parametrized by the energy as the sole relevant quantity.
Then, we have identified two asymptotically-relevant regimes,
depending on the particle energy:
(i)~For energy below the percolation threshold, any particle
is trapped in a finite-size region,
so that $\langle r^2 \rangle \rightarrow \textrm{const}$\ (classical localization regime), 
along the $r=x,y$ direction.
The probability of diffusion
(defined as the average density probability of finding an atom at a given distance
from its origin point)
develops exponentially decaying tails,
which correspond to the probability distribution of the size of
the classically-localized regions.
Due to the pure topographical nature of 
the latter, the probability of diffusion
is anisotropic with the same anisotropy factor as the speckle potential.
(ii)~For energy above the percolation threshold, 
there appear infinite-size allowed regions in which the gas expands diffusively,
$\langle r^2\rangle \propto t$ (normal diffusion regime).
The probability of diffusion is a Gaussian function 
with anisotropy factor (along the $x$ and $y$ directions) 
significantly different from the 
anisotropic factor of the speckle potential.
The diffusion coefficients along the two main directions of
the speckle potential strongly depend on the particle energy.
More precisely, we found a power-law scaling,
$D_r \sim E^\gamma$
with $\gamma \simeq 3.15$ for a blue-detuned speckle potential
and $\gamma \simeq 3.45$ for a red-detuned speckle potential.

In addition, a transient sub-diffusion regime,
where $\langle r^2\rangle \propto t^{\alpha_r}$ with $\alpha_r<1$,
characterizes intermediate time scales, relevant to recent experiments~\cite{Robert_2010}.
This regime appears below and slightly above the percolation threshold.
It is interpreted as due to the dynamics of exploration of
finite-size allowed regions which are disconnected from each other or
connected by narrow bottlenecks.
For energy below the percolation threshold, the long-time
dynamics crosses over to classical localization ($\alpha_r = 0$).
Conversely, above the percolation threshold, it crosses over to
normal diffusion ($\alpha_r = 1$).
This example shows that subdiffusion should not be regarded as
precursor of classical localization.

Using the above results as a groundwork,
we have studied the expansion dynamics of
a cloud of cold atoms with broad energy distributions,
as relevant to recent experiments.
First, we have calculated the energy distribution of the cloud 
in the presence of disorder. 
Then, we have discussed the dynamics. 
The time-dependent profiles are in general anisotropic and non-Gaussian,
with tails 
governed by diffusive atoms.
We have proposed a method to extract the energy-dependent diffusion
coefficients from these tails, which involve many energy components.
The experiments of Ref.~\cite{Robert_2010} are realized in a quasi-2D geometry,
which does not significantly change the above-discussed dynamics.
The energy-dependent diffusion coefficients 
extracted from experimental data
are in fair agreement with numerical calculations performed in the
precise experimental geometry.

In the future, it would be interesting to experimentally study the
classical localization regime as discussed in the paper.
In particular, working at lower temperature, it would be possible 
to create a gas of atoms with a significant fraction below the percolation
threshold. Then, studying the fraction of classically localized atoms
versus temperature, for instance, would give direct access to the percolation
transition and, in particular, to the critical exponent.
Another interesting direction would be to extend the study of transport
in anisotropic 2D speckle potentials in the quantum regime,
where traces of weak and strong localization effects can be searched for.

\vspace{1cm}
\textbf{Acknowledgements}\\
\\*
We thank Boris Shapiro for useful comments on the manuscript
and M. Besbes. This research was supported by
CNRS,
CNES as part of the ICE project,
Direction G\'en\'erale de l'Armement,
the European Research Council (FP7/2007-2013 Grant Agreement No.\ 256294)
Agence Nationale de la Recherche (Contract No.\ ANR-08-blan-0016-01),
RTRA Triangle de la Physique,
IXSEA,
EuroQuasar program of the EU,
and MAP program SAI of the European Space Agency (ESA).
LCFIO is member of Institut Francilien de Recherche sur les Atomes Froids (IFRAF).
We acknowledge the use of the computing facility cluster GMPCS of the LUMAT
federation (FR LUMAT 2764).

\vspace{1cm}
\textbf{References}\\
\\*



\begin{thebibliography}{99}
\bibitem{nota_continuity_speckle} By numerical construction, 
the 2D speckle potential is continuous at the box boundaries.
\bibitem{note_Boltzmann} Alternatively, one can introduce the ballistic path length,
$l_r^*(E) \equiv \tau_r^*(E) \sqrt{2E/m}$,
which is the characteristic length traveled by a particle,
beyond which its trajectory is affected by the disorder.
\bibitem{note_redspeack}
For a red-detuned speckle potential ($\Vr<0$), we find:
\be
n(E_\textrm{p}) = \frac{e^{-E_\textrm{p}/\Vr }}{\vert \Vr \vert} \left\{ \begin{array}{rl}
 & \me^{E/\Vr } \, \Theta(E-E_\textrm{p})  \mbox{ for $E\leq 0$} \\
 & \Theta(E_\textrm{p}/\Vr)  \qquad \quad \mbox{ for $E\geq 0$}.\end{array} \right. \nonumber
\ee
For $E \leq 0$, the average values of the potential energy and of kinetic energy along each direction
are $\langle E_\textrm{p} \rangle = \Vr + E $
and $\langle E_\textrm{k} \rangle = \vert \Vr \vert$, respectively.
For $E \geq 0$, we find
$\langle E_\textrm{p} \rangle = \Vr$ and $\langle E_\textrm{k} \rangle = E + \vert \Vr \vert$.
The latter relations are intuitive in the limit $E \gg \vert \Vr \vert$.
The fact that they hold for $E\geq 0$ is a consequence of phase-space ergodicity in the 2D geometry.
\bibitem{note_AnisoFact}
In the numerics, we find a slow dependence of
the anisotropy factor of the diffusion, $D_x/D_y$, versus the energy $E$.
However, in the whole energy window that is considered here,
we find $D_x/D_y \simeq 3.1$ with an accuracy better than $10\%$.
\bibitem{note_NEred} For a red-detuned speckle potential ($\Vr<0$), Eq.~(\ref{<nE1>})
is given by
\be
\langle n(E) \rangle = \frac{\beta}{1+\beta \vert \Vr \vert }
\left\{ \begin{array}{rl}
 & e^{+E/\vert \Vr \vert} \quad \mbox{ for $E\leq 0$} \\
 & e^{-\beta E} \qquad \,\,\,\, \mbox{ for $E\geq 0$},\end{array} \right. \nonumber
\ee
while Eq.~(\ref{<nE2>}) is 
\be
\langle n(E) \rangle = \beta(1-\beta \vert \Vr \vert)
\left\{ \begin{array}{rl}
 & e^{-E(\beta -1/\vert \Vr \vert)} \mbox{ for $E\leq 0$} \\
 & e^{-\beta E} \qquad \quad \,\,\,\, \mbox{ for $E\geq 0$}.\end{array} \right. \nonumber
\ee
\bibitem{note_ballistic} For an isotropic, ballistic expansion we have
$f_E(\vecr, t; \vecr_0, t_0) = \Theta(t-t_0)\, \delta\big[ \vert \vecr - \vecr_0 \vert -
(t-t_0) \sqrt{2 E / m} \big]/ 2 \pi (t-t_0) \sqrt{2 E / m}) $.
\bibitem{note_barE} For the distribution of Eq.~(\ref{NE1}) we have 
$\bar E = k_\textrm{B} T +\Vr$. For the parameters of Fig.~(\ref{Figure:BECMSF}), 
$k_\textrm{B} T = \Vr$, we have $\bar E=2\Vr$ and $\tau^*(\bar E)/\tau_R \sim 10$.
\bibitem{note3Deqs} In the 3D case, the classical equations of motion read
\beq
\frac{\ud \tilde{\vecr}}{\ud \tilde{t}} & = & \tilde{\vecp}, \nonumber \\
\frac{\ud \tilde{\vecp}}{\ud \tilde{t}} & = & - \tilde{\nabla} \left[\upsilon_{\mathrm{iso}} (\tilde{x} \sin\theta - \tilde{z}\cos\theta,\tilde{y}) + \frac{(\omega_z \taur)^2 \tilde{z}^2}{2} \right], \nonumber
\eeq
where the reduced variables are defined in Sec.~\ref{EquationMotion}.
\end{thebibliography}
\end{document}